# What Factors Control Shale Gas Production and Production Decline Trend in Fractured Systems: A Comprehensive Analysis and Investigation

HanYi Wang

Petroleum and Geosystem Engineering Department, The University of Texas at Austin, USA

**Summary**

One of the biggest practical problems with the optimization of shale gas stimulation design is estimating post-fracture production rate, production decline, and ultimate recovery. Without a realistic prediction of the production decline trend resulting from a given completion and reservoir properties, it is impossible to evaluate the economic viability of producing natural gas from shale plays.

Traditionally, decline curve analysis (DCA) is commonly used to predict gas production and its decline trend to determine the estimated ultimate recovery (EUR), but its analysis cannot be used to analyze what factors influence the production decline trend due to lack of underlying support of physics, which make it difficult to guide completion designs or optimize field development.

In this article, we presented a unified shale gas reservoir model, which incorporates real gas transport, nano-flow mechanisms and geomechanics into fractured shale system. This model is used to predict shale gas production under different reservoir scenarios and investigate what factors control its decline trend. The results and analysis presented in the article provide us a better understanding of gas production and decline mechanisms in a shale gas well with certain conditions of the reservoir characteristics. More in-depth knowledge regarding the effects of factors controlling the behavior of the gas production can help us develop more reliable models to forecast shale gas decline trend and ultimate recovery. This article also reveals that some commonly hold beliefs may sound reasonable to infer production decline trend, but may not be true in a coupled reservoir system in reality.

**Introduction**

With ever increasing demanding for cleaner energy, unconventional gas reservoirs are expected to play a vital role in satisfying the global needs for gas in the future. The major component of unconventional gas reservoirs comprises of shale gas. Shales and silts are the most abundant sedimentary rocks in the earth's crust and it is evident from the recent year's activities in shale gas plays that in the future shale gas will constitute the largest component in gas production globally, as conventional reservoirs continue declines (Energy Information Administration 2015). According to GSGI, there are more than 688 shales worldwide in 142 basins and 48 major shale basins are located in 32 countries (Newell 2011). Better reservoir knowledge and advancing horizontal drilling and hydraulic fracturing technologies make the production of shale gas resources economically viable and more efficient.

Unlike conventional reservoirs, the shale gas reservoirs tend to be more expensive to develop and require special technologies to enable the gas to be produced at an economical rate due to its extremely low matrix permeability and porosity. Thus modeling the shale gas production and its declines correctly is essential to predict how fast the gas can be produced and turned into revenue from each well, and the economic viability of producing natural gas from the operated shale plays. **Fig.1** shows the average gas production rate per well, for wells that grouped by their first production date in different shale basin across the United States. It can be observed that gas production rate declines rapidly within the first a few months of production. We can also notice that the peak of production rate does not occur in the first month, this is because early time production are affected by wellbore clean-up effects, when gas are produced with water and fracturing fluids.

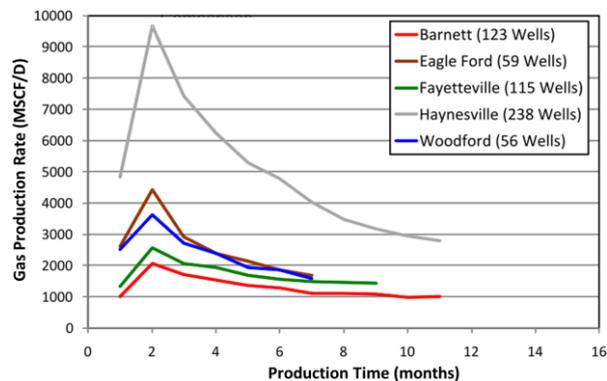

**Fig. 1  Gas production rate in different shale gas basins (Modified from Baihly et al. 2010)**

The estimated ultimate recovery (EUR) of shale gas wells has been forecasted in a number of ways within the industry, primarily based on initial production rates (IPs) and decline trends. The decline curve analysis (DCA) is probably the most frequently used production forecasting tool for shale gas reservoirs due to its relative simplicity and speed. The common methods used to estimate oil and gas reserves rely on a set of empirical production decline curves based on the following hyperbolic function (Arps 1945):

$$q_t = q_i(1 + bD_i t)^{-\frac{1}{b}}$$

where $q_t$ is the production rate at time t, $q_i$ is the initial production rate at time t = 0, $D_i$ and $b$ are two constants (the former is the initial rate of decline in production and $b$ is the rate of change in $D_i$ over time, which control the curvature of the decline trend).

The Arps equation was designed for conventional reservoirs where the boundary-dominated flow is the norm. However, shale gas reservoirs are characterized by transient production behavior and in general boundary-dominated flow only occurs in later times. Shale gas production rates generally have steep initial decline trends and gradually flatten out (also shown in Fig. 1). The resulting DCA can often yield b values greater than 1.0. This not only gives higher EUR values, but also leads to an extrapolation of infinite EUR. When this occurs, production decline curve must be converted to another type of function at some point so as to restrict the EUR to be finite. In addition, the Arps model has too few freely adjustable parameters to fit the data fully, so biased and subjective extrapolation could be off the accepted range (e.g., during early well productions, models with vastly different b values all look similar). The differences only show up in the long-term, leading to vastly different EUR values.

The flaws in Arps model has led to the development of many new DCA models for predicting estimated EUR in the shale gas wells, such as the power law exponential model (Ilk et al. 2008), logistic growth analysis (Clark et al. 2011) and Duong's model (Duong 2011), etc. Even though all these models were formulated differently, they are all empirical equations and lack of underlying support of physics, so the same production data may lead to different estimation of EUR and production decline trend, with different practice of tuning parameters. Moreover, shale has very distinct characteristics even it lays next to each other in the same area, and decline rate may differ from stage to stage, well to well (King 2010), thus make it impossible to extrapolate the results of DCA model from one well to another.

In general, few shale wells have gone through whole life cycles, and there are significant uncertainties that associated with different decline models when trying to estimate the volume of technically recoverable reserves. Since production decline most dramatically in the first a few years and the economically preferred completion designs may be more driven by the net present value derived in the first 5 years of production rather than the ultimate recovery of the well (Barree et al. 2015). This early 5-year period that represents most of the useful economic life of the well, is hard to predict with enough confidence using empirical models due to short of production data. In addition, empirical models cannot be used to analyze what factors cause the shift of production decline curve in field cases with different practices, which makes it impossible to assign value to one design over another and equally impossible to optimize the treatment for whichever goal is sought, either acceleration of recovery or increase in reserves.

A better understanding of gas production and its decline trend in a shale gas well, in particular, regions with certain conditions of the reservoir characteristics, will give a better insight of whether the DCA model used is appropriate. More in-depth knowledge regarding the effects of factors controlling the behavior of the production decline can help us develop more reliable models to forecast shale gas production, decline trend and ultimate recovery. In this study, a unified shale gas reservoir model is used to investigate what factors control shale gas production and its decline trend in a systematic manner. The results of this study provide us a comprehensive understanding of the correlations between various factors and production decline trend in hydraulic fractured, horizontal gas wells.

The structure of this article is as follows. First, some unique characteristics that influence shale gas production and decline trend are reviewed and discussed. Next, a unified reservoir model is used to predict the long-term well performance of hydraulically fractured wells under different reservoir in-situ conditions. Then, the simulation results are analyzed and compared to examine what factors mostly control shale gas production and decline trend. Finally, conclusion remarks and discussions are presented. The proposed mathematical framework for unified shale gas reservoir modeling, which incorporates real gas transport, nano-flow mechanisms and geomechanics into fractured shale system, is presented in **Appendix**.

## Shale Gas Production and Physical Mechanisms

In this section, some important mechanisms that may control shale gas production and decline trend are presented and discussed, those factors and their coupled effects on the overall horizontal gas well performance will be examined extensively throughout this article.

### Adsorption Gas

In shale gas reservoirs, gas can be stored as compressed fluid inside the pores or it can be adsorbed by the solid matrix. The gas adsorption in shale-gas system is primarily controlled by the presence of organic matter and the gas adsorption capacity depends on TOC (Total Organic Carbon), organic matter type, thermal maturity and clay minerals. Generally, the higher the TOC content, the greater the gas adsorption capacity. In addition, a large number of nanopores lead to significant nanoporosity in shale formations, which increases the gas adsorption surface area substantially (Javadpour et al. 2007). The amount of adsorbed gas varies from 35-58% (Barnett Shale, USA) up to 60-85% (Lewis Shale, USA) of total gas initial in-place

(Darishchev et al. 2013). **Fig. 2** shows laboratory measurement of gas adsorption capacity from a shale sample at different temperatures. It can be observed that the reservoir pressure must be sufficiently low to liberate the adsorbed gas. For organically rich shales, the ultimately recoverable amount of gas is largely a function of the adsorbed gas that can be released (desorbed). Because most adsorbed gas can only be released at low reservoir pressure, due to the extremely low permeability of shale matrix, even with hydraulic fracturing, there is still significant amount of residual absorption gas that cannot be produced, unless additional stimulation methods are used, such as thermal stimulation (Wang et al. 2014; Wang et al. 2015a; Yue et al. 2015). Understanding the effects that initial adsorption, and moreover, desorption has on gas production and decline trend will increase the effectiveness of reservoir management and economic evaluations.

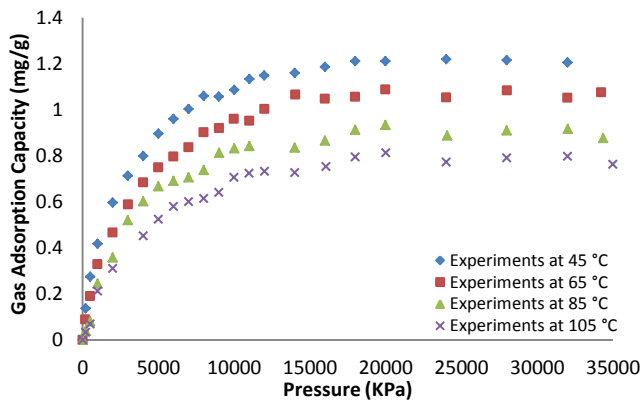

**Fig. 2** Gas adsorption capacity from Jiashiba Shale sample at different temperatures (Yue et al. 2015)

**Matrix Apparent Permeability and Its Evolution**

Darcy's equation, which models pressure-driven viscous flow, works properly for reservoirs where continuum theory holds and fluid velocity can be assumed to be zero at the pore wall (Sherman 1969). However, in shale reservoirs which have pore throat radii in the range of nanometers, fluid continuum theory breaks down and gas molecules follow a somewhat random path while still maintaining a general flow direction governed by the pressure gradient. Molecules strike against the pore walls and tend to slip at pore walls instead of having zero velocity. In order to propose corrections for non-Darcy flow over different flow regions in nanopore space, numerous authors (Civan 2010; Darabi et al. 2012; Fathi et al. 2012; Sakhaee-Pour and Bryant 2012; Singh et al. 2014) have quantified these effects by modifying the slippage factor or determining the apparent permeability as a function of Knudsen number.

Even though numerous theoretical and empirical models have been developed for apparent permeability enhancement in nanopore structures, to account for the effects of non-Darcy flow/gas-slippage behavior, they ignore the changes in pore structure geometry induced by stress variations and the release of the adsorption gas layer during production. The assumption of constant pore radius is valid under steady-state flow, however, within these ranges of ultralow permeability, transient and pseudo-steady-state flow dominate the entire production life of shale formations. With constant decreasing pore pressure and increasing effective stress, the formation microstructure will be affected. Wang and Marongiu-Porcu (2015) presented a comprehensive literature review on the evolution of matrix permeability during pressure depletion and proposed a unified matrix permeability model which incorporates the mechanisms of non-Darcy flow/Gas-Slippage, the release of the adsorption gas layer and geomechanical effects into a coherent global model, as shown in **Fig. 3**. Their works indicate that despite rock compaction, the apparent permeability in shale matrix increases during production due to the combined effects of non-Darcy flow/gas-slippage and release of the adsorption gas layer. The evolution of matrix permeability as reservoir pore pressure declines can also make a difference in determining gas production rate. In this article, a more general formula for global shale matrix permeability evolution is implemented in the model and will be presented in **Appendix**.

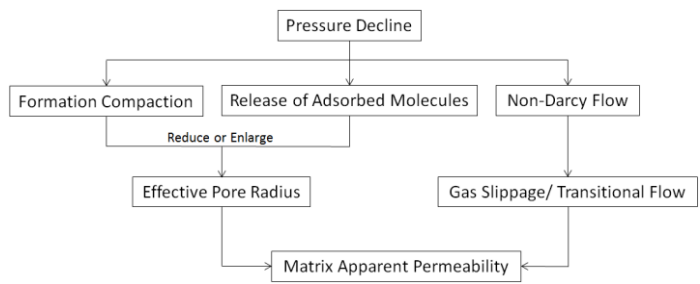

**Fig. 3** Mechanisms that alter shale matrix apparent permeability during production (Wang and Marongiu-Porcu 2015)

**Fracture Networks and Pressure/Stress-Dependent Conductivity**

Many field experiments demonstrated that a propagating hydraulic fracture encountering natural fractures may lead to the arrest of fracture propagation, fluid flow into natural fracture, the creation of multiple fractures and fracture offsets. Under some specific stress conditions and the perturbations of hydraulic fracturing, the natural fractures may coalesce to form very conductive, larger scale channels and completely change the formation rock flow properties. Now it is a well-known fact that most brittle shale formations have massive pre-existing natural fracture networks that are generally sealed by precipitated materials weakly bonded with mineralization. Such poorly sealed natural fractures can interact heavily with the hydraulic fractures during the injection treatments, serving as preferential paths for the growth of complex fracture network (Cipolla et al. 2011; Zakhour et al. 2015). However, the permeability/conductivity of the fracture networks (include both propped and un-propped fractures) is a function of confining stress, fluid and proppant type for a specific reservoir (Fredd et al. 2000; Ghassemi and Suarez-Rivera 2012). Laboratory experiment can be used to determine the expected reduction in fracture conductivity caused by the increasing effective stress by using correlations (Cho et al. 2013). The reduction of fracture conductivity due to proppant/asperities crash or embedment, when fluid pressure declines in the fracture system, can strongly influence the efficiency of the fracture networks and gas production rate.

**Non-Stimulated Reservoir Volume**

The interaction between pre-existing natural fractures and the advancing hydraulic fracture is a key issue leading to complex fracture patterns. Large populations of natural fractures are sealed by precipitated cement which is weakly bonded with mineralization that even if there is no porosity in the sealed fractures, they may still serve as weak paths for the growing hydraulic fractures. Even if the proppant material cannot reach a certain branch of the network, or even if the branch is not connected to the perforations, the slippage caused residual opening or some other type of "rock failure" provides increased flow capacity within its vicinity. The observations of fracture complexity have led to the introduction of Stimulated Reservoir Volume (SRV), which is used to describe the shape and size of a fracture network created by hydraulic fracturing in low-permeability reservoirs. SRV is considered as an important parameter for well performance in low permeability reservoirs because their productivity depends primarily on enhanced permeability of the SRV and the reservoir area contacted by the SRV (Jackson et al. 2013; Palmer et al. 2014; Suliman et al. 2013).

When the fracture spacing design is not optimized (e.g., a large number of tightly spaced perforation clusters are simultaneously propagating fractures), the stress interference may prohibit some fractures from growing (Shin and Sharma 2014). Field study (Minner et al. 2003) also indicates that 80% fracture volume created at the heel and toe of a horizontal well and only 20% fracture volume created at the mid-lateral, due to poor stimulation design. If rock property is heterogeneous along the horizontal wellbore, fractures may only grow in brittle sections and leave the ductile sections unstimulated, because lower pressure is needed to initiate and propagate in brittle rocks (Wang 2015; Wang 2016b; Wang et al. 2016). Sub-optimal hydraulic fracture design and reservoir heterogeneity may lead to large reservoir volume unstimulated. The reservoir volumes, that are not overlapped by SRV and remain its original flow capacity, will be classified and investigated as Non-SRV in the following sections.

**Numerical Modeling and Base Case Simulation**

In order to investigate the impact of various mechanisms on shale gas production process, a unified shale gas reservoir simulator is developed. The fully coupled physical process implemented involves fluid flow within the formation matrix and fractures, shale gas adsorption and desorption, real gas properties and in-situ stress that changes dynamically during production. This state of the art simulator enables us to investigate what factors have the most influential impact on the global well performance and ultimate recovery in shale gas formations, by combining real-gas nano-flow mechanisms and geomechanical effects in fracture systems. For easy interpretation and generality of analysis, wellbore clean-up process will not be modeled and discussed in this study.

Even though it is possible to simulate a complete section of a horizontal well with multiple transverse fractures, it is more efficient to simulate a unit stimulated reservoir volume (SRV) and apply symmetric boundary conditions along the boundaries, and then the total gas production from a horizontal well can be determined by adding the contributions from each SRV unit. **Fig. 4** shows the plane view of SRV unit with a hydraulic fracture that penetrates into the formation from the horizontal well bore and intercepts by abundant natural fractures (the red line depicts horizontal wellbore, the blue line and the randomly distributed line segments represent the hydraulic fracture and natural fractures respectively). Constant production pressure is applied at the wellbore and no-flux conditions are applied at SRV boundaries.

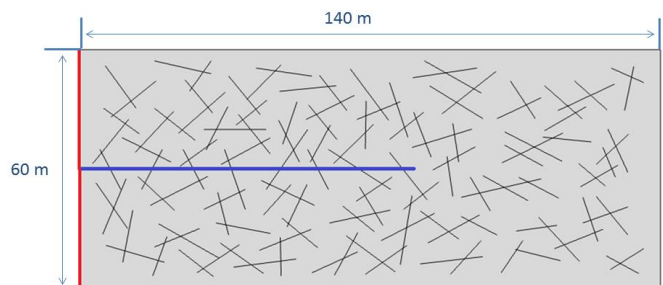

**Fig. 4 A plane view of SRV containing hydraulic fracture and natural fractures**

In practice, the primary hydraulic fracture and secondary fracture networks within each SRV unit, can be created statistically by using seismic, well and core data (Ahmed et al. 2013; Cornette et al. 2012 ) and by matching hydraulic fracture propagation models. Hydraulic fracture models that commonly used in industry relies on linear elastic fracture mechanics (LEFM) (Johri and Zoback 2013; Weng 2014), but it is only valid for brittle rocks (Wang et al. 2016), recent studies demonstrate that using energy based cohesive zone model can resolve the issue of singularity at the fracture tip and these cohesive zone based hydraulic fracture models not only can be used in both brittle and ductile formations, but also can capture complex fracture evolution, such as natural fracture interactions (Guo et al. 2015), fracture reorientation from different perforation angles (Wang 2015), producing well interference (Wang 2016a), the effects of fracturing spacing and sequencing on fracture interaction/coalescence from single and multiple horizontal wells (Wang 2016b) and integrate microseismic data with non-planar fracture modeling (Haddad et al. 2016). Combine hydraulic fracture modeling and microseismic data, the spatial distribution of fracture networks can be estimated or calibrated by various methods (Aimene and Ouenes, 2015; Huang et al. 2014; Johri and Zoback 2013; Yu et al. 2014). And the properties of the propped fracture and un-propped fracture can be determined through conductivity test under various pressures and confining stresses (Fredd et al. 2000; Ghassemi and Suarez-Rivera 2012), so that the parameters define transport capacity of fractures can be determined (e.g., fitting Eq.A.8 with laboratory data). With all these information, the spatial and intrinsic properties of DFNs can be estimated. How to map fracture networks across different fracturing stages based on microseismic data/hydraulic fracture modeling and fully characterize the entire horizontal completion is beyond the scope of this article. Because natural fractures are formed within geological time scale and the formation itself may have gone through multiple tectonic events, the distribution and orientation of the natural fractures do not necessarily relate to the current in-situ stresses. The distribution and orientation of natural fractures are randomly placed inside SRV in this study. We first specify the density of DFNs (how many DFNs we need in our given SRV), then the center points of each DFN is arbitrarily generated in the SRV, and a random DFN orientation and DFN length (within predefine range) are assigned to each center point to generate two end points of DFN, connect these two points and all the DFNs are created in the simulation domain. The final distribution of natural fractures is pictured in Fig. 4.

**Table 1** shows all the input parameters used for the base case simulation, which includes reservoir conditions, drainage geometry, fracture conductivity, real gas properties, and gas adsorption parameters. All these input parameters are reasonably designed to reflect a typical shale reservoir formation, where around 60% of initial gas-in-place comes from adsorbed gas, this can be the case in many organic-rich shale formations (Darishchev et al., 2013). The impact adsorption gas on total material balance and gas production will be examined later. For the base case simulation, the evolutions of matrix and fracture permeability are disabled (matrix and fracture permeability maintain constant during production) and their effects on gas well performance will be investigated separately the following sections.

Table 1    Input Parameters for Base Case Simulation

| Input Parameters | Value |
|---|---|
| Initial reservoir pressure, $p_i$ | 35[MPa] |
| Wellbore pressure, $p_{wf}$ | 5[MPa] |
| Initial matrix permeability, $k_{m0}$ | 100[nd] |
| Initial hydraulic fracture permeability, $k_{f0}$ | $10^4$[md] |
| Initial natural fracture permeability, $k_{nf0}$ | 200[md] |
| Initial matrix porosity, $\phi_{m0}$ | 0.01 |
| Initial average pore throat radius, $r_0$ | 5[nm] |
| Material constant for porosity change, $C_\emptyset$ | 0.035 |
| Hydraulic fracture porosity, $\phi_f$ | 0.5 |
| Natural fracture porosity, $\phi_{nf}$ | 0.01 |
| Hydraulic fracture width, $d_f$ | 0.01[m] |
| Natural fracture width, $d_{nf}$ | 0.0001[m] |
| Fracturing spacing, $Y_e$ | 60[m] |
| Reservoir thickness, $H$ | 20[m] |
| Drainage length parallel to the hydraulic fracture, $X_e$ | 140[m] |
| Half fracture length, $x_f$ | 80[m] |
| Total number of drainage units, $n$ | 20 |
| Reservoir temperature, $T$ | 350[K] |
| Density of formation rock, $\rho_m$ | 2600 [kg /$m^3$] |
| Langmuir volume constant, $V_L$ | 0.0017[$m^3$/kg] |
| Langmuir pressure constant, $P_L$ | 10[MPa] |
| Diameter of adsorption gas molecules, $d_m$ | 0.414[nm] |
| Average molecular weight, $M$ | 16.04[g/mol] |
| Critical temperature of mix gas, $T_c$ | 191 [K] |
| Critical pressure of mix gas, $P_c$ | 4.64[MPa] |
| Horizontal minimum stress, $S_h$ | 40[MPa] |
| Horizontal maximum stress, $S_H$ | 45[MPa] |
| Overburden stress, $S_v$ | 50[MPa] |
| Biot's constant, $\alpha$ | 1 |
| Young's modulus, $E$ | 25[GPa] |
| Poisson's ratio, $v$ | 0.25 |

**Fig. 5** shows the pressure distribution in the simulated SRV unit after 1 and 12 months of production. It can be observed that the pressure transient behavior has reached the boundary after one month of production, owing to the existence of abundant and well-connected conductive fracture networks, and pseudo-steady-state flow has already ensued from the initial transient flow, to dominate the rest of the production history. After 12 months of production, most of the area that penetrated by the main hydraulic fracture has been well depleted due to the well-connected fracture system and effective linear flow from matrix to fracture, but the area in front the main hydraulic fracture tip still remain largely untapped.

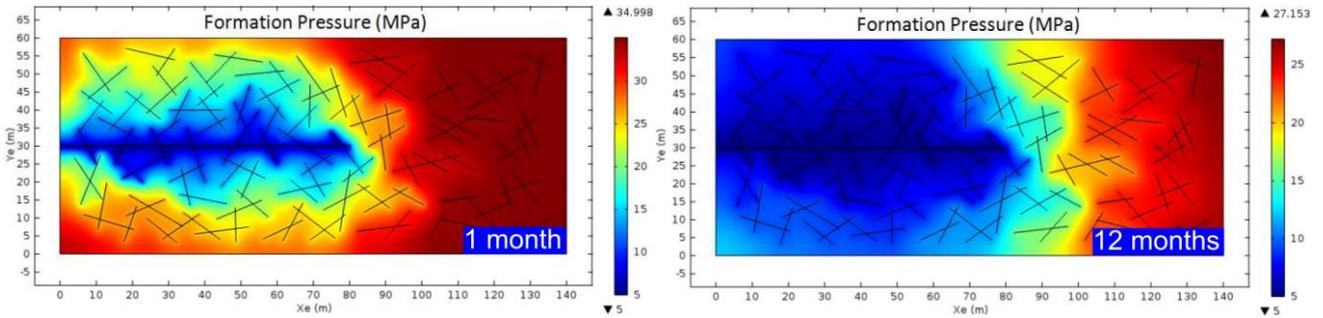

**Fig. 5 Pressure distribution in the SRV after 1 and 12 months of production**

**Fig. 6** shows the gas viscosity, gas density, fraction of adsorption gas produced and mean effective stress profiles in the simulated SRV after 1 month of production, all of which are time-space dependent as pore pressure declines. It can be observed that gas viscosity and gas density are lower in the region that adjacent to the hydraulic fracture, where pore pressure is also lower. And the highest mean effects stress also concentrates around the low-pressure zone. The lower the pore pressure, the higher the effective stress. It can be further noticed that, at most, around 53% of adsorption gas can be produced in the region that close to hydraulic fracture surface with current operation conditions (i.e., 5 MPa wellbore pressure) and adsorption gas properties ( determined by Langmuir volume and pressure constant). This implies that there are still a large amount of residual adsorption gas cannot be ever recovered no matter how long it takes, unless lower bottom-hole pressure or other stimulation methods are applied.

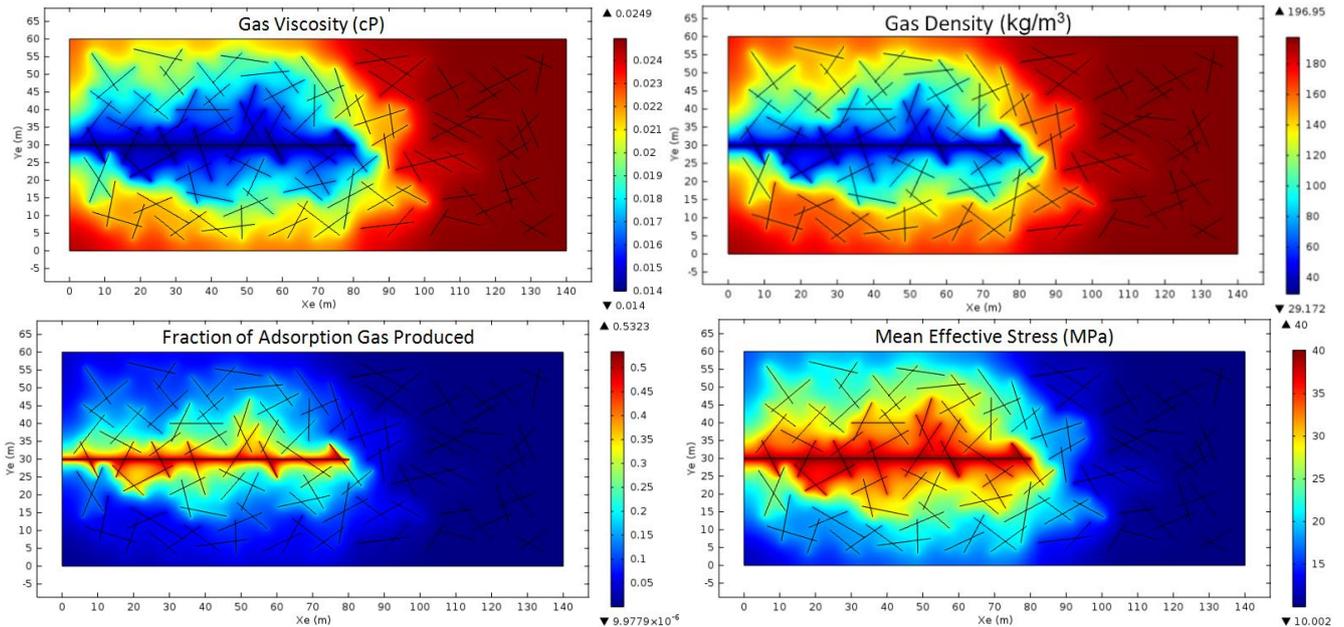

**Fig. 6 The distribution of gas viscosity, gas density, fraction of adsorption gas produced and mean effective stress in the SRV after 1 month of production**

**Fig. 7** shows pressure profile along the hydraulic fracture after 1 and 12 months of production. The hydraulic fracture half-length is 80 meters (on the horizontal axis, 0 and 80 represent the wellbore and fracture tip, respectively). It is obvious that pressure drop does exist along the fracture length, but the magnitude of pressure loss is insignificant, and becomes even smaller as time goes on. This is due to the fact that the matrix permeability is extremely small and turbulence flow is unlikely to happen under such low level of flow rate, and the permeability of the fracture is large enough. In late times, when the pressure drop is even smaller along the fracture path, the hydraulic fracture can be viewed as infinite conductivity.

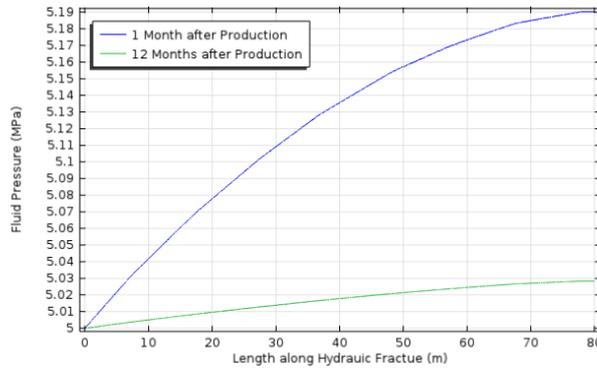

**Fig. 7 Fluid pressure distribution along hydraulic fracture length**

**Fig. 8** shows the cumulative production and average monthly production rate. It clearly demonstrates that the production rate declines dramatically in the first two years, and becomes more stable in later times. **Fig. 9** shows the monthly production decline rate and normalized production rate (monthly production rate divided by production rate for the first month). It can be observed that the gas production rate declines at 20%-40% per month in the first a few months and the decline rate stabilizes around 5% per month after 20 months of production. It can be also noticed that the monthly production rate declines 90% at the end of the second year of production, which is a very typical scenario for unconventional reservoirs.

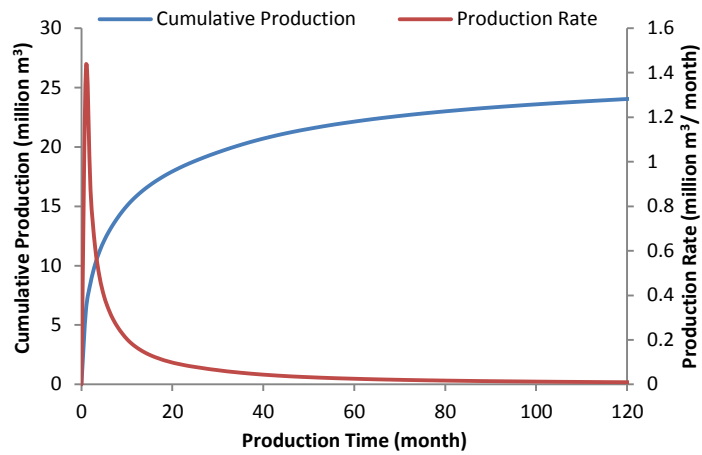

**Fig. 8 cumulative production and production rate**

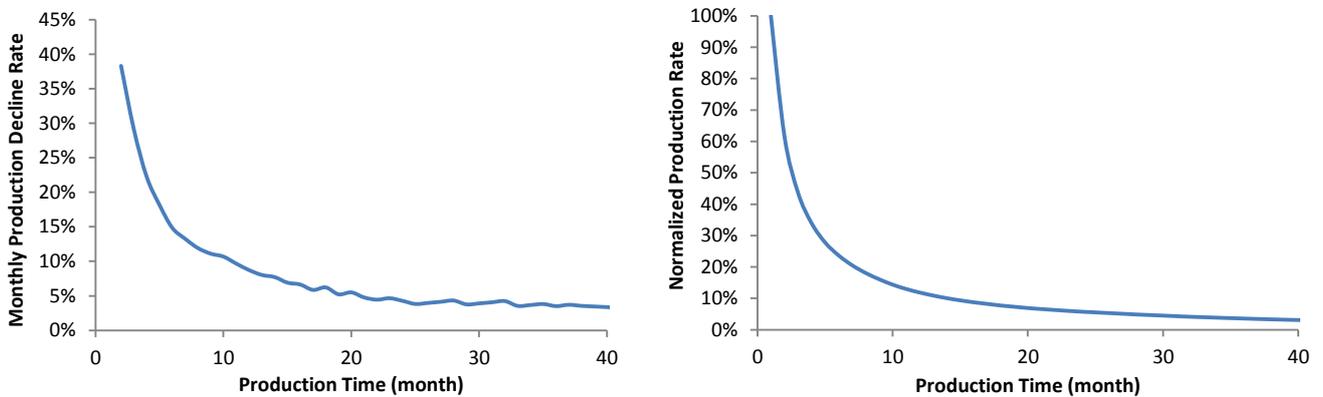

**Fig. 9 Monthly production decline rate and normalized production rate**

### Impact of Adsorption Gas

In the above base case simulation, around 60% gas initially in place is adsorption gas. In the section, we remove all the adsorbed gas and leave only free gas in the porous medium, to examine its impact on gas production and decline trend. **Fig. 10** shows the pressure distribution in the simulated SRV unit after 1 and 12 months of production. Compared with Fig. 5, where adsorption gas is included, the results indicate that pressure declines more rapidly and a larger area in the simulated SRV can be depleted with the same production time.

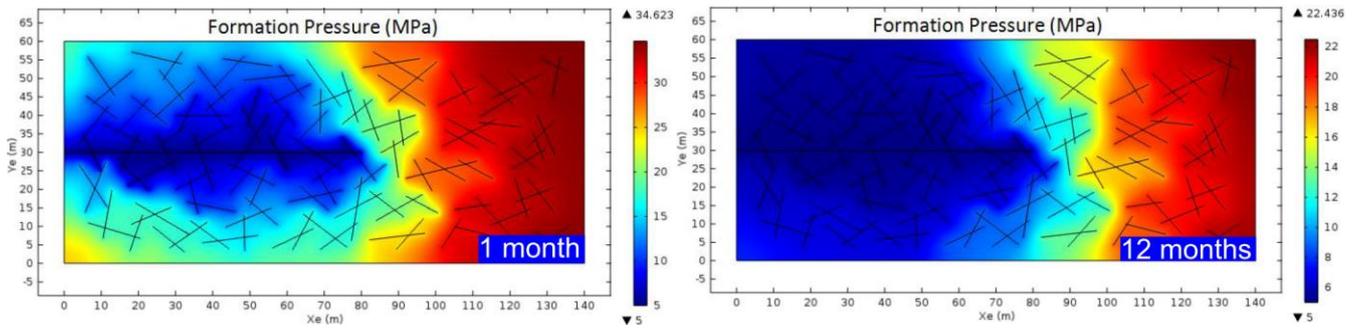

**Fig. 10 Pressure distribution in the SRV after 1 and 12 months of production, without adsorption gas**

**Fig. 11** demonstrates that more gas can be produced when adsorption gas is included in the model, because there is much less total initial gas in place if only consider free gas alone. **Fig. 12** indicates that the existence of adsorption gas indeed reduces production decline rate, especially in the early stages. This can be explained by the fact that the release of adsorption gas can compensate pressure loss in the SRV to some degree, so more gas can be produced at the same pressure level. However, in the later stage, the impact of adsorption gas on pressure declines becomes less obvious, this is because most of the recoverable adsorption gas have been produced as the overall pressure declines in the SRV, and the remaining adsorption gas contribute much less portion of total gas produced, as the pressure declines further. It is contrary to the common belief that adsorption gas becomes important only when the average pressure in the reservoir drops to a certain level during the state of pseudo-state flow. In fact, adsorption gas has more impact in early transient flow stages when the desorbed gas continues compensating pressure loss inside the matrix that is depleting by the adjacent low-pressure sinks (i.e., well-connected fracture networks). Combine Fig.11 and Fig.12, we can conclude that adsorption gas has more impact on cumulative production (as governed by Eq.A.2 and A3, from a material balance perspective) than it has on long-term production decline trend.

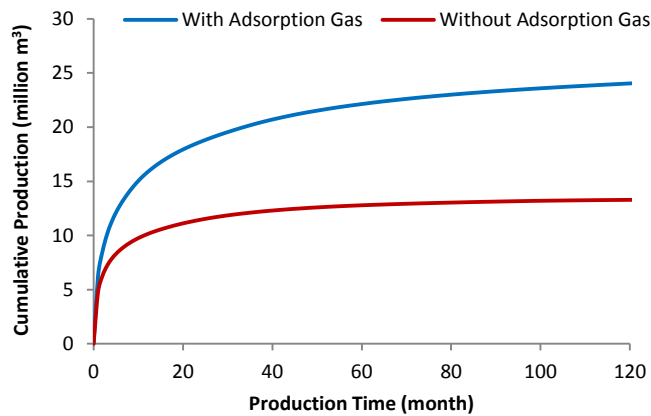

**Fig. 11 Cumulative production with and without adsorption gas**

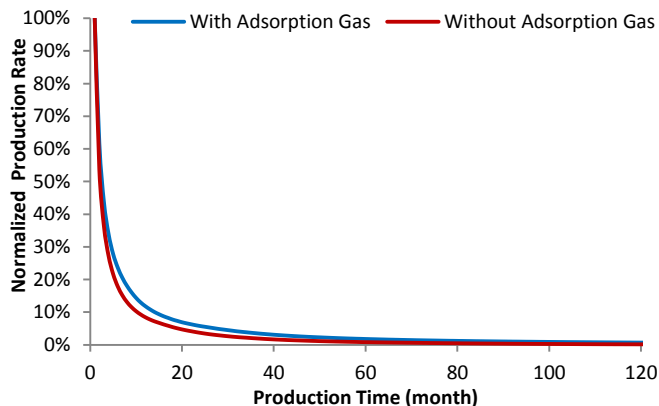

**Fig. 12 Normalized production rate with and without adsorption gas**

### Impact of Matrix Permeability

Next, the impact of matrix permeability and its evolution during production are investigated. First, the matrix permeability is modified to be 10 nd and 500 nd from the base case simulation (matrix permeability is 100 nd). **Fig. 13** shows the pressure profiles in the SRV after 1 month of production with different matrix permeability. It can be clearly noticed that when matrix permeability is only 10 nd, pressure depletion is constrained to the close vicinity of fracture networks that well-connected the hydraulic fracture. However, when the matrix permeability increases to 500 nd, the hydraulic fracture becomes the main pressure sink source and the role of natural fractures become less significant.

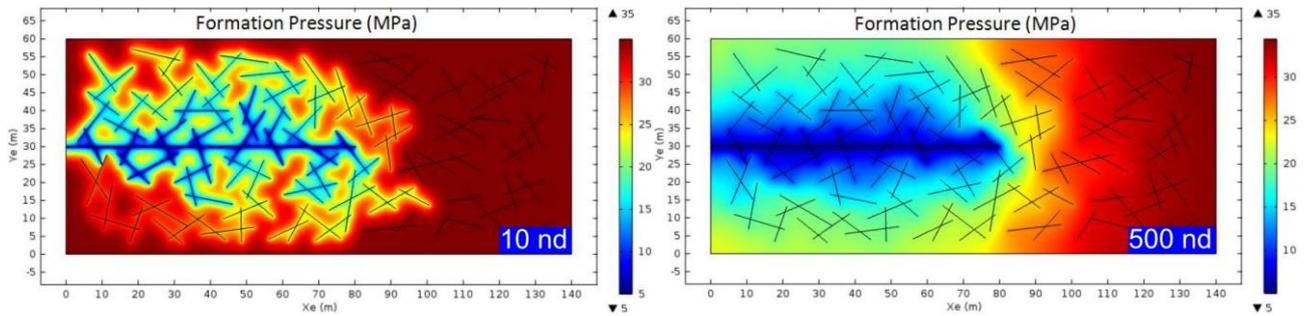

**Fig. 13 Pressure distribution in the SRV after 1 month of production with different matrix permeability**

As mentioned in the previous section, the apparent permeability in the matrix of shale gas formations can be altered by non-Darcy flow, gas slippage, the release of adsorption layer and rock compaction during production. A unified shale matrix permeability model (Eq.A.27) that encompasses all these mechanisms is implemented in the simulator. When the matrix permeability is allowed to evolve as pore pressure declines, the in-situ matrix permeability will be space and time dependent, as opposed to maintaining its initial matrix permeability. **Fig. 14** shows the values of Knudsen Number and matrix apparent permeability in the SRV after 1 month of production. The Knudsen number is a dimensionless parameter that can be used to differentiate flow regimes in conduits at micro and nanoscale, and Darcy's law is only valid when Knudsen number is less than $10^{-3}$ (Knudsen 1909). The simulation results indicate that Knudsen Number increases as pore pressure declines, and the value of Knudsen Number reveals that slip and transition flow regimes are most likely to be encountered. It can be also observed that the matrix apparent permeability is the highest (around 227 nd) in the low-pressure zone region, while remaining the same as initial matrix permeability (100 nd) in the region that has not been tapped by pressure depletion. The local matrix permeability will evolve as pressure transient front edge continues to propagate: starting from areas that are close to pressure sink to the rest of the SRV.

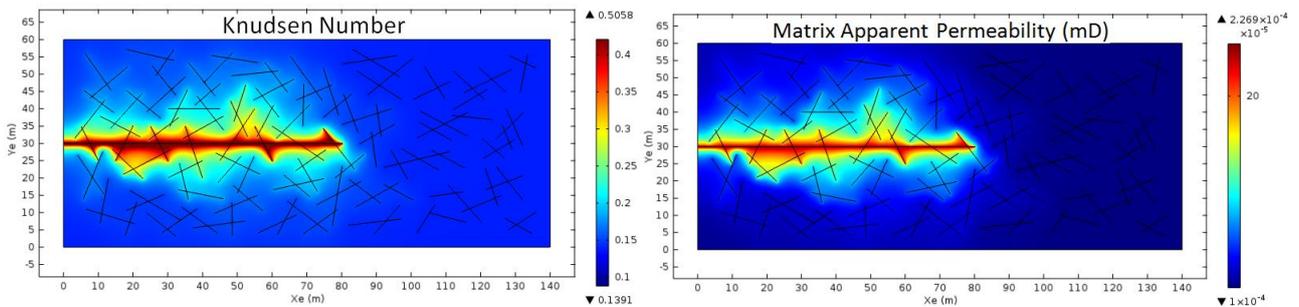

**Fig. 14 Knudsen Number and matrix apparent permeability profiles in the SRV after 1 month of production, Km=100 nd**

**Fig. 15** shows the cumulative production of different matrix permeability. The results demonstrate that more gas can be produced with higher matrix permeability, and the evolution of matrix permeability during production can indeed make a difference in gas production, due to the enhanced local matrix apparent permeability. **Fig. 16** shows the normalized production with different permeability. It indicates that, even though, higher matrix permeability contributes to higher production rate, it also increases production decline rate.

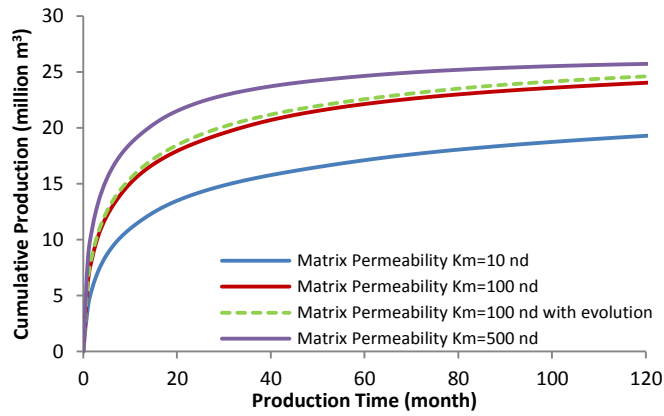

Fig. 15 Cumulative production with different matrix permeability

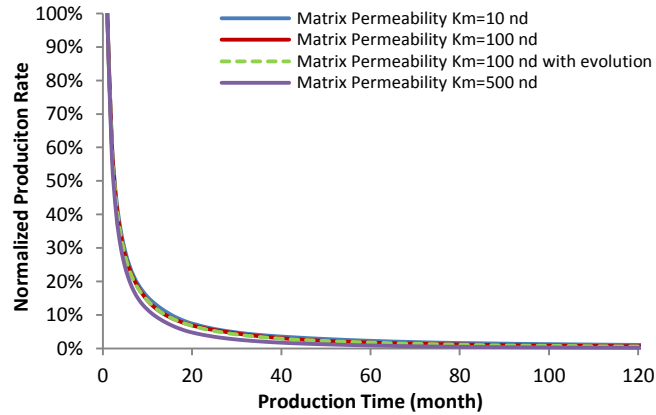

Fig. 16 Normalized production rate with different matrix permeability

It seems the impact of matrix permeability evolution during production has limited impact on overall gas production. It is possible that with abundant, conductive fracture networks, the general the role of matrix permeability itself is diminished. Next, all the natural fractures are removed from the simulation domain and the hydraulic fracture serves as the only conductive path for gas flow. **Fig. 17** and **Fig. 18** show the cumulative production and normalized production rate with and without matrix permeability evolution. The results demonstrate that matrix permeability evolution has a more pronounced impact on cumulative production if fracture networks are not included, but it has a negligible impact on production decline trend.

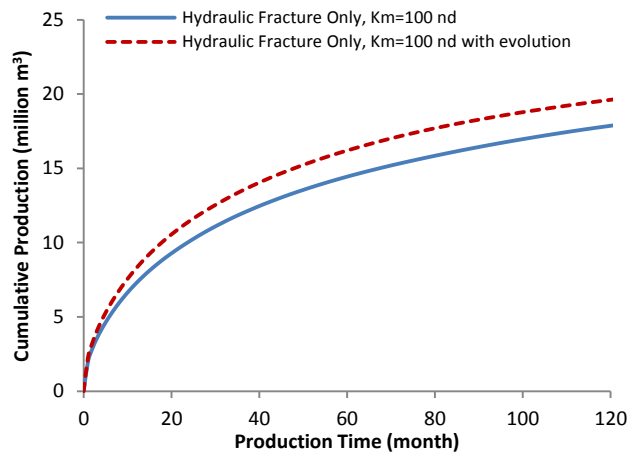

Fig. 17 Cumulative production with and without matrix permeability evolution

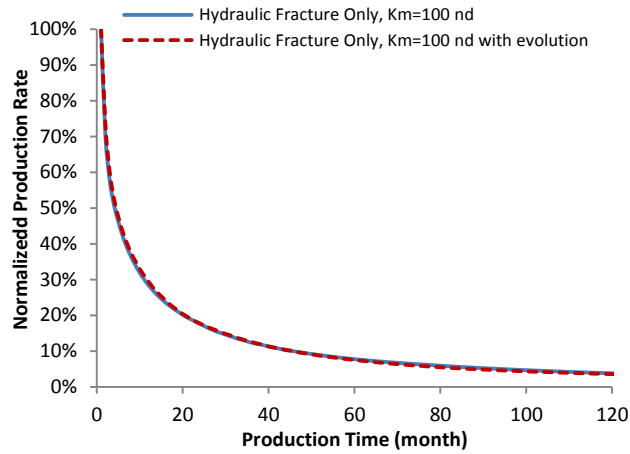
**Fig. 18 Normalized production rate with and without matrix permeability evolution**

### Impact of Non-SRV

As discussed earlier, sub-optimal completion design (stress interference) and reservoir heterogeneity (brittle and ductile rocks) can lead to unevenly stimulated reservoir volume along the horizontal wellbore. In order to investigate the impact of Non-SRV, where the pre-existing natural fractures have not been activated by the stimulation job, the simulation domain is extended beyond the SRV regime, as shown in **Fig. 19**. All the input parameters and boundary conditions remain the same as the base case.

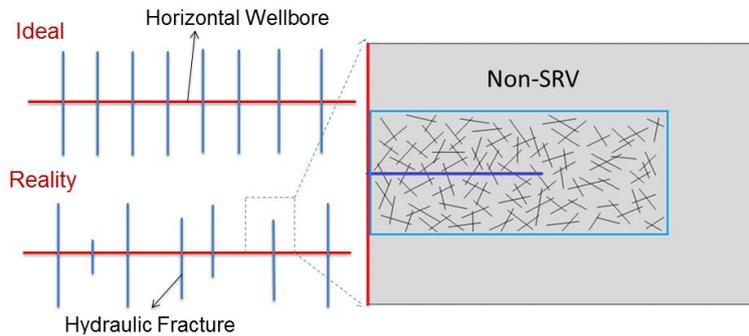
**Fig. 19 A plane view of multiple transverse fractures and simulation domain that contain Non-SRV**

**Fig. 20** shows the pressure profiles in the simulated domain after 1 and 12 months of production. It can be observed that after 1 month of production, the front edge of transient flow regime has reached the boundary of SRV, and after 12 months of production, most area within the SRV, that penetrated by the hydraulic fracture, has been well depleted. But the pressure in the Non-SRV regime still remains relatively high.

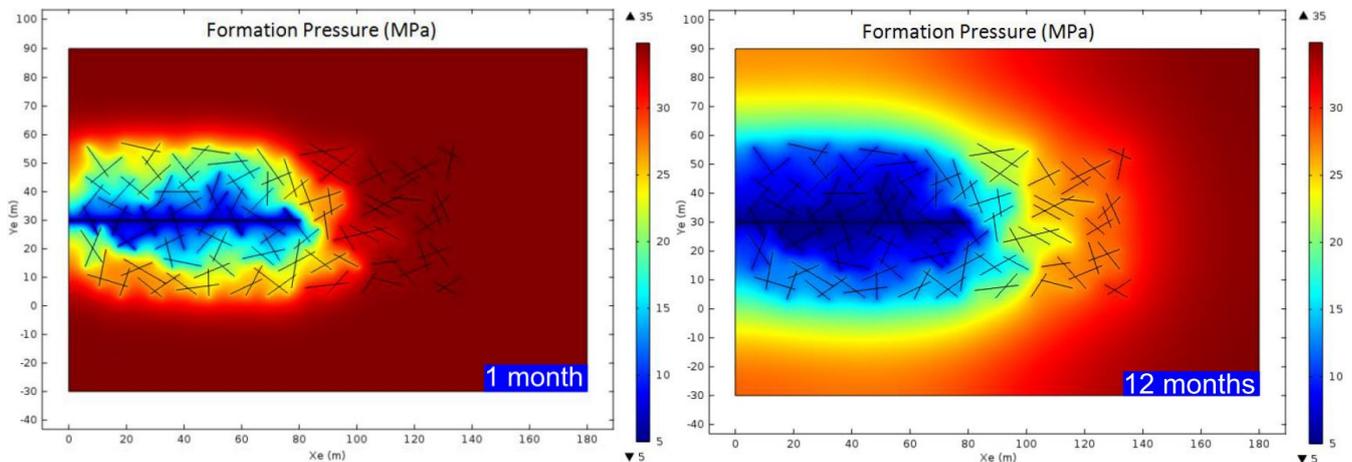
**Fig. 20 Pressure distribution in the simulation domain after 1 and 12 months of production with dense fracture networks inside the SRV**

**Fig. 21** shows the impact of Non-SRV on cumulative production with different matrix permeability. The results indicate the matrix permeability have a vital role in determining the influence of Non-SRV. When matrix permeability is high (e.g., 100 nd), less time is needed to observe the effects of Non-SRV and much more gas can be produced from Non-SRV. On the contrary, when matrix permeability is small (e.g., 10 nd), the effects of Non-SRV on gas production is restricted. The smaller the matrix permeability, the lesser impact that Non-SRV has on production. The results also demonstrate the existence of Non-SRV amplifies the effects on matrix evolution. It should be mentioned that accumulative production assumed the same number of unit production pattern. If the length of a horizontal well is fixed, larger fracture spacing, with the inclusion of Non-SRV, shall have less production, due to less number of unit production pattern.

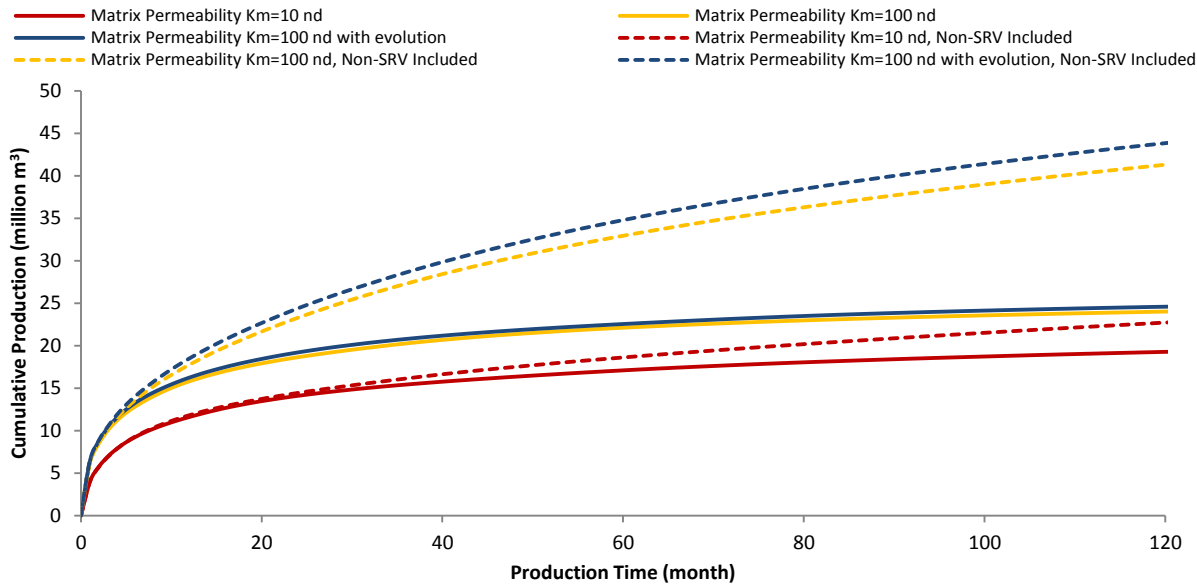

**Fig. 21 Cumulative production with and without Non-SRV under different matrix permeability**

**Fig. 22** shows the impact of Non-SRV on production decline with different matrix permeability. The results indicate that the gas production decline rate can be mitigated if Non-SRV exists. And the matrix permeability plays a different role in production decline trend, compared with the case where only SRV is considered: The higher the matrix permeability, the lower the decline rate, because pressure loss in the SRV can be compensated more quickly by the surrounding Non-SRV. This is contradictory to the effect of matrix permeability on decline trend that is shown in Fig. 16, where only SRV is included.

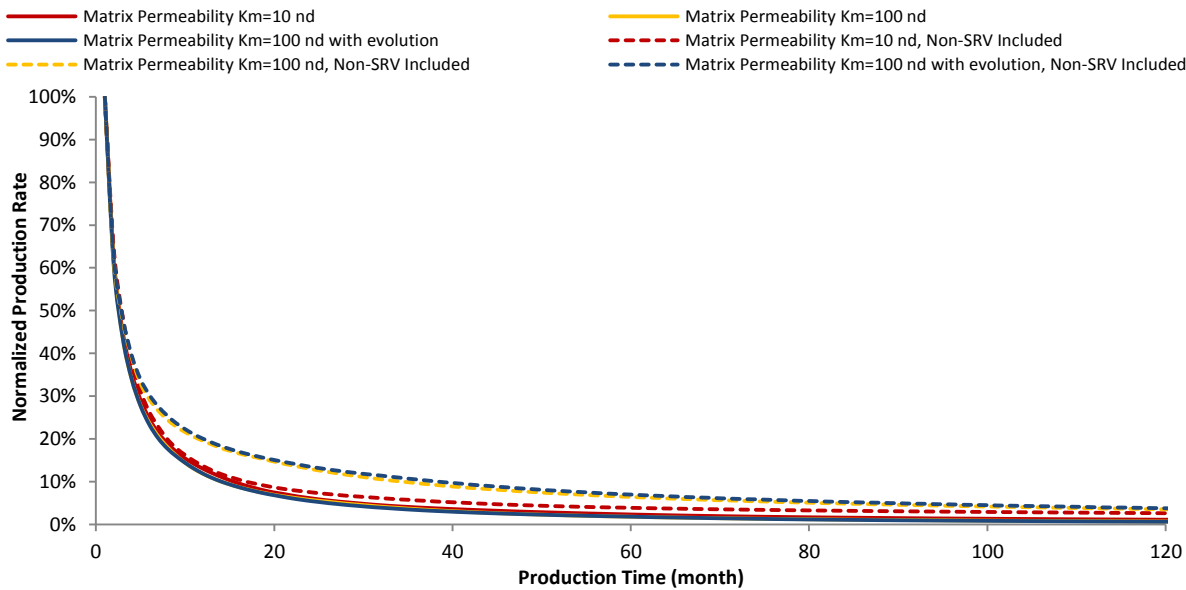

**Fig. 22 Normalized production rate with and without Non-SRV under different matrix permeability**

Next, we investigate effects of the density of conductive natural fractures inside the SRV, by reducing the number of natural fractures from previous simulation case. **Fig. 23** shows the pressure profiles in the simulated domain after 1 and 12 months of production when the fracture network in the SRV is sparse. Compared with Fig. 20, where the density of natural fractures is much higher, fewer reservoir volumes are depleted during the same production period.

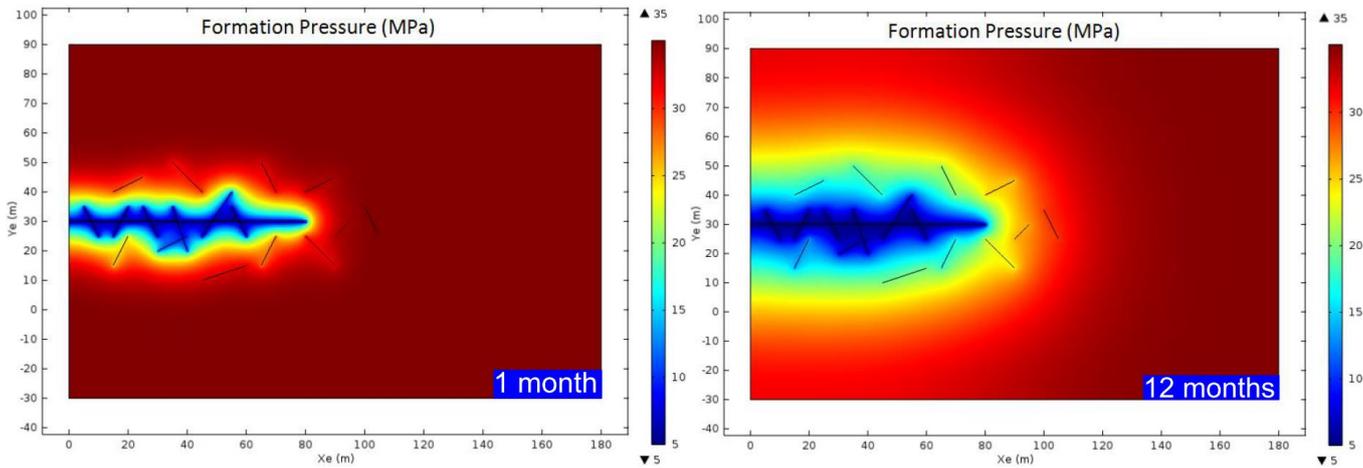

**Fig. 23 Pressure distribution in the simulation domain after 1 and 12 months of production with sparse fracture networks inside the SRV**

**Fig. 24** and **Fig. 25** demonstrate the impact of fracture network density in the SRV on cumulative production and production decline trend. No surprising that dense fracture networks lead to higher production rate and higher cumulative production, but it also results in more rapid production decline rate.

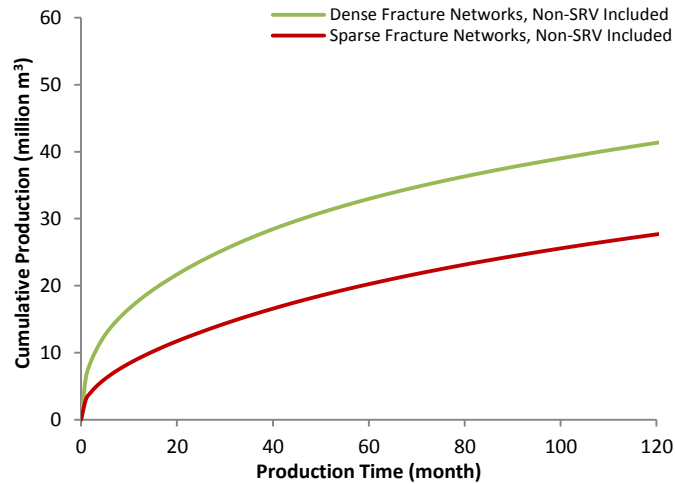

**Fig. 24 Cumulative production with Non-SRV under different density of natural fractures**

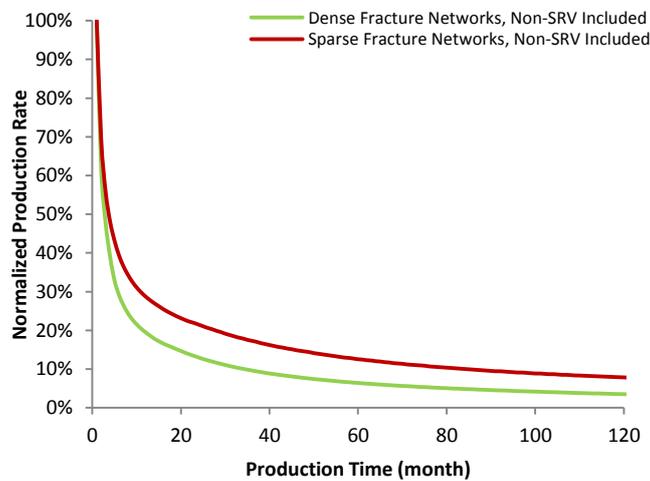

**Fig. 25 Normalized production rate with Non-SRV under different density of natural fractures**

## Impact of Fracture Networks

In this section, the role of natural fractures, their distribution and pressure/stress dependent conductivity on gas production and production decline trend will be extensively examined. **Fig. 26** shows the relationships between fluid pressure and fracture permeability that will be implemented as pressure-dependent-permeability for fractures in the following simulation. As can be seen, when the fluid pressure declines from initial pore pressure (35 MPa) to bottom-hole pressure (5MPa), the permeability of hydraulic fracture reduced to 78% of its initial value, while the permeability of natural fractures reduced to 10% of its initial value. Ideally, laboratory experiment should be used to determine the propped and un-propped fracture conductivity as a function of effective stress (Fredd et al. 2000; Ghassemi and Suarez-Rivera 2012), and in the reservoir, local effective stress increases as pore pressure decreases during depletion. For the purpose of simplicity and easy interpret results, the local relationship between fracture conductivity and pore pressure is represented by the relationship between fracture permeability and pore pressure, by assuming the fracture width remains the same. Because fracture conductivity is the product of fracture permeability and fracture width, shift fracture permeability has the same effect as shift fracture conductivity on simulation results. In addition, the fracture storage effects have negligible impact on gas production, because the driving forces of primary shale gas recovery come from the gas volume expansion and gas desorption with declining pressure. The conception of pressure-dependent-permeability for fractures, can sometimes be referred to as pressure-dependent-conductivity, stress-dependent-permeability or stress-dependent-conductivity in literature. For consistency, the term "pressure-dependent-permeability (PDP)", will be used throughout this article, to represent the changes in fracture conductivity as local effective stress increases and fluid pressure declines.

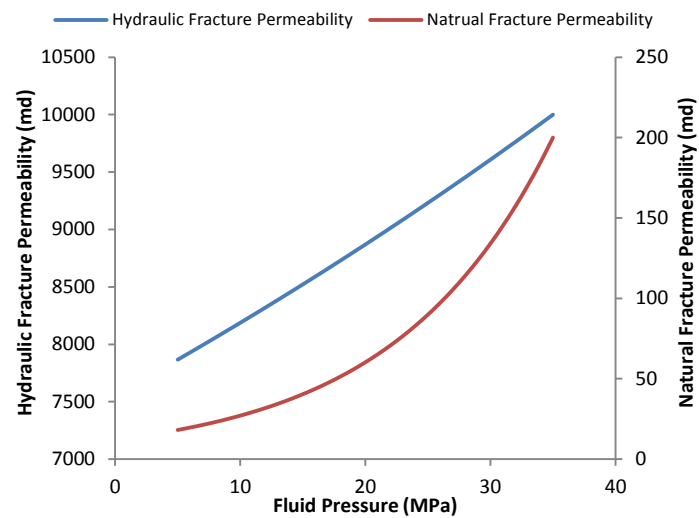

**Fig. 26 Pressure dependent fracture permeability**

**Fig. 27** shows the cumulative production under different scenarios. In order to have a more comprehensive comparison, the cases of a vertical well and no fracture networks are included. The "Dense Fracture Networks" represents the fracture network topology in the base case simulation (as shown in Fig. 5), while the "Sparse Fracture Networks" represent much less dense fracture networks in the SRV (as shown in Fig. 23). The results indicate that the denser the fracture networks, the larger the fracture surface area, the higher the gas production. The pressure dependent fracture permeability indeed reduced the production by impairing the efficiency of fracture conductivity on fluid transport. **Fig. 28** shows the corresponding production decline trend. Combined with Fig. 27, it can be concluded that even though, dense fracture networks with large fracture surface area lead to higher absolute production rate and cumulative production, but they also result in higher production decline rate.

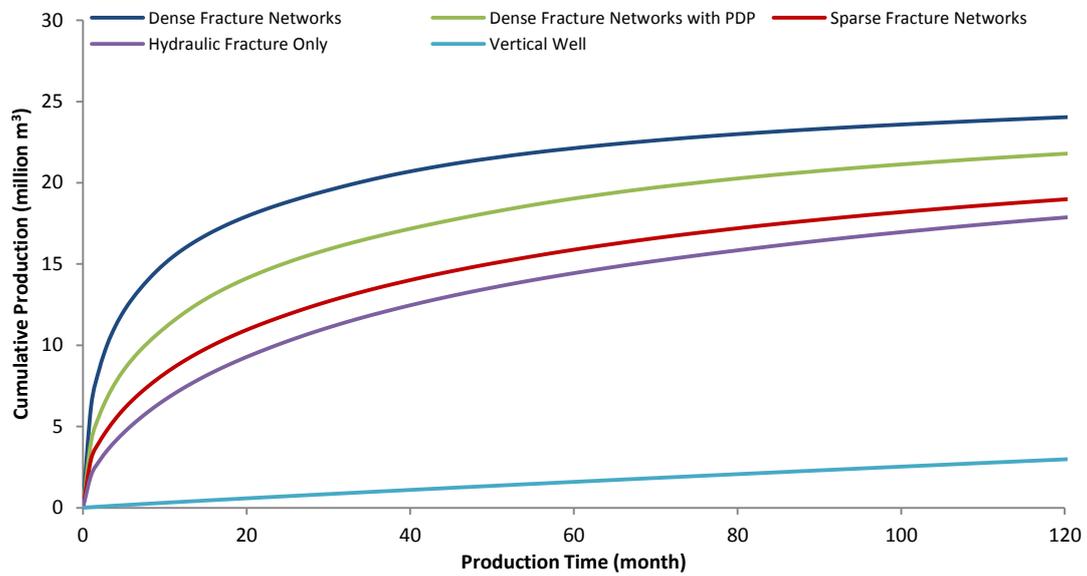

Fig. 27 Cumulative production under different scenarios, Km=100 nd

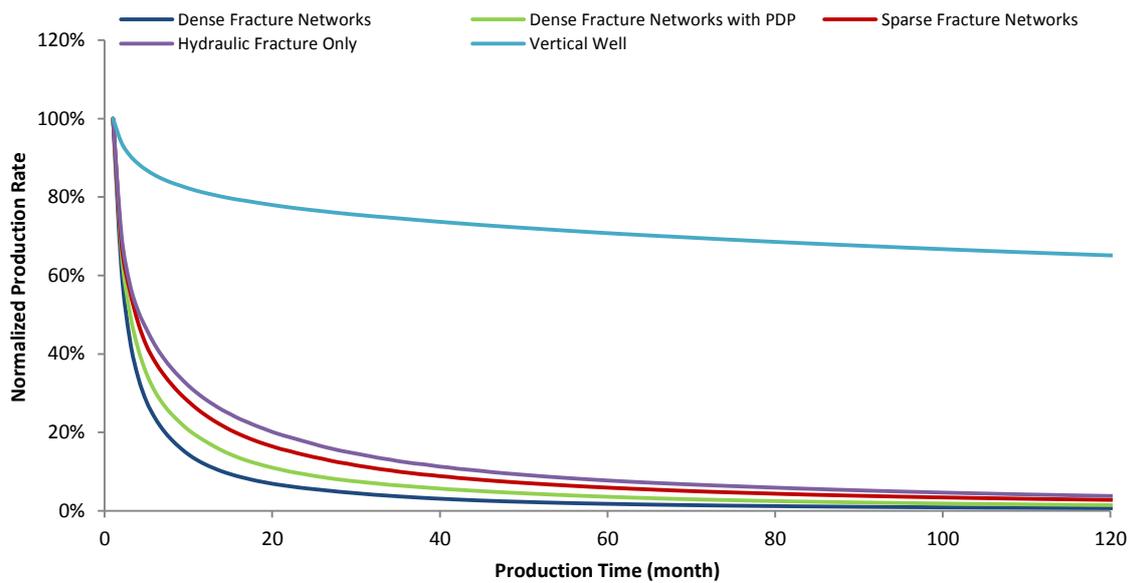

Fig. 28 Normalized production rate under different scenarios, Km=100 nd

Next, the matrix permeability is reduced to 10 nd, **Fig. 29** and **Fig. 30** shows the impact of matrix permeability on gas production and its decline trend in different scenarios. Compared with previous results, it can be observed that the existence of fracture networks has more impact on gas production when matrix permeability is lower (i.e., the cumulative production in the case of dense fracture networks is more than 2 times of that with sparse fracture networks when matrix permeability is 10 nd, while the cumulative production in the case of dense fracture networks is only around a quarter more than that sparse fracture networks when matrix permeability is 100 nd). It can be also noticed that regardless of matrix permeability, denser fracture networks and larger fracture surface area inside SRV always lead to higher production and higher decline rate over time.

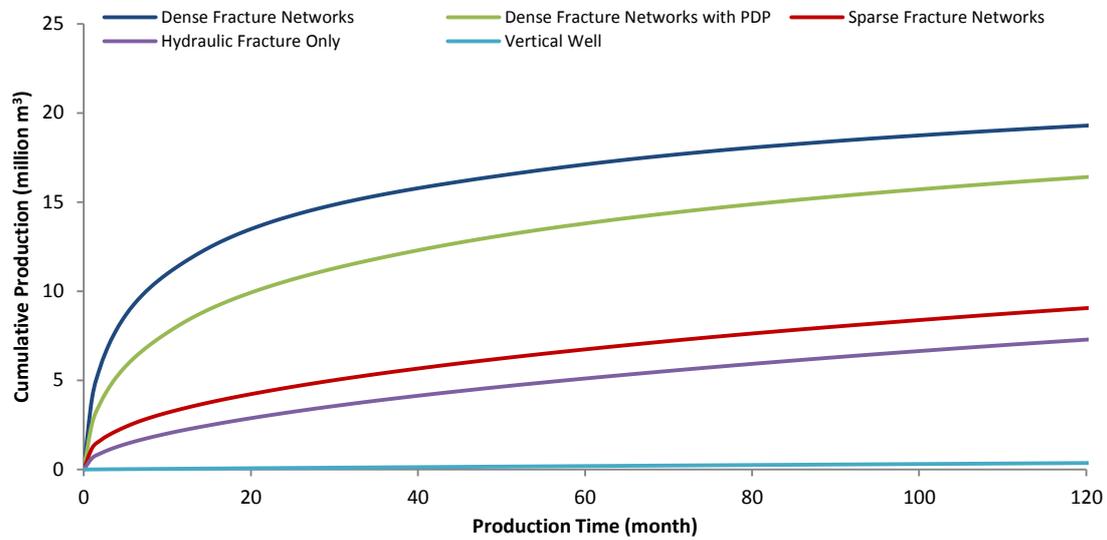

Fig. 29 Cumulative production under different scenarios, Km=10 nd

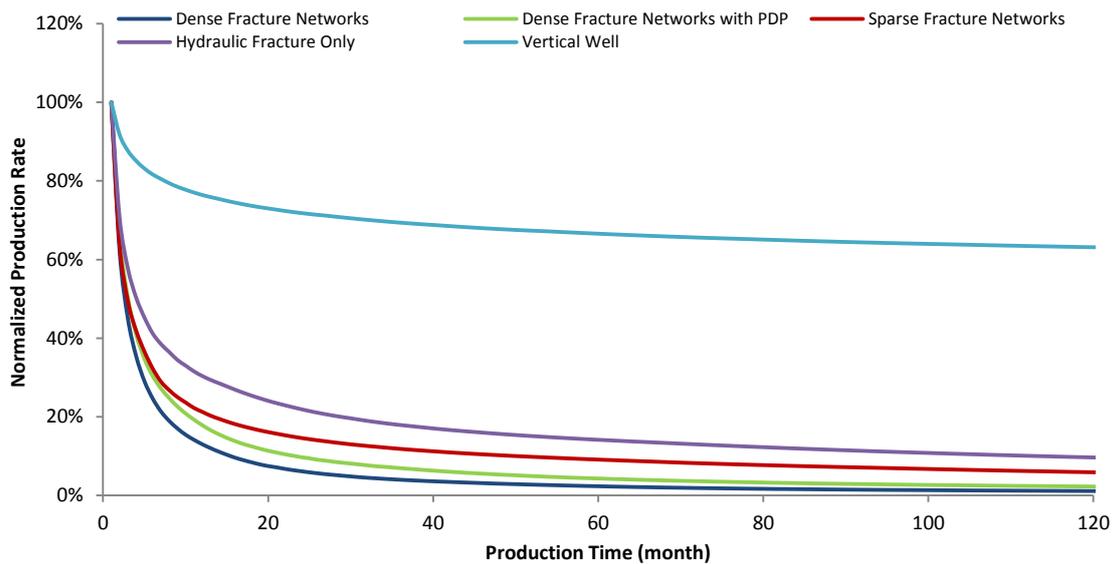

Fig. 30 Normalized production rate under different scenarios, Km=10 nd

Previous study (Baihly et al. 2010) claims that the production decline trend in Barnett Shales is more gradual than other shale plays, mainly because the natural fracture set in Barnett Shales are more abundant, better connected and the conductivity is less pressure dependent. However, our above analysis leads to opposite conclusions. More adsorption gas and the inclusion of Non-SRV can reduce the production decline rate, but a higher density of natural fracture networks with better conductivity always lead to steeper production decline trend. This can be explained by simple reasoning: with more natural fractures that connected to the hydraulic fracture, more gas can be produced initially due to more reservoir contact area. But as the shale matrix that close to the fracture surface depleted, it is more difficult to deplete the gas inside shale matrix that at a distance away from the fracture surface, so the relative production rate declines more rapidly with larger fracture surface area with higher initial production rate. At the same time, higher production rate lead to sooner termination of transient flow inside SRV and the ensued steady-steady, boundary dominated flow makes the production rate declines further.

It is intuitive to think that fracture dependent permeability/conductivity reduce overall production and increase production decline rate. Baihly et al. (2010) also reported that the Haynesville and Eagle Ford Shales have steeper decline rates because of the degradation of fracture network conductivity during production. However, the results from our above study indicate that pressure dependent fracture permeability does impair the overall production, but it actually reduces production decline rate, regardless of matrix permeability. If we re-examine Fig. 7, it can be noticed that the pressure drop along the path of fracture is small and if the all the conductive fractures are well connected initially, once a low-pressure sink is imposed at the wellbore or perforation tunnel, the pressure transient behavior can dissipate very quickly and the fluid pressure inside all the well-connected fracture networks drops to the level of wellbore pressure. This means the reduction of fracture conductivity (due to increased effective stress) happens once production starts, instead of occurring gradually as reservoir pore pressure declines. **Fig. 31** shows the impact of pressure dependent fracture permeability on gas production in the first month, it is clear that the reduction of fracture permeability reduces initial production rate. Even tough pressure dependent fracture permeability reduces production decline rate, it also reduces absolute production rate at the very beginning, and that it why it harms the overall cumulative production.

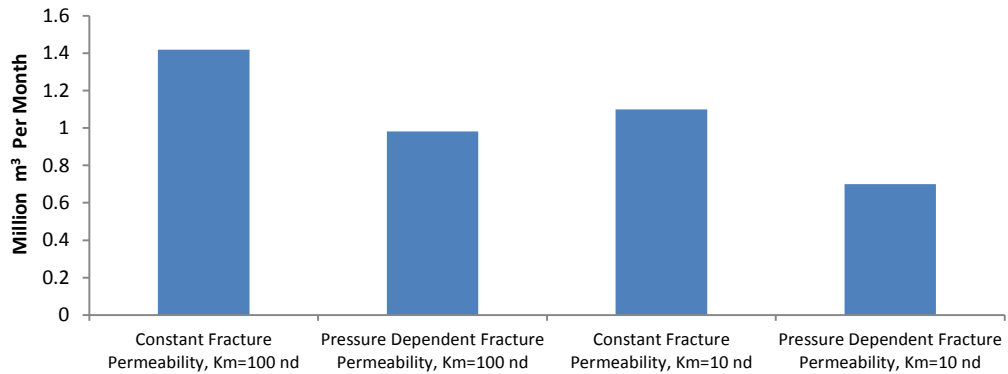

**Fig. 31 The impact of fracture PDP on the first-month production rate with dense fracture networks**

It seems that effects of fracture PDP on production decline trend contradict common beliefs at this point. Next, the natural fractures in the SRV are re-organized, such that they are poorly connected and do not intercept by hydraulic fracture. Then the low-pressure sink along the hydraulic fracture will not impact the conductivity of natural fractures at the beginning of production. **Fig. 32** shows the pressure distribution inside SRV after 1 and 12 months of production with poorly connected natural fractures. Compared with Fig. 5, where the fracture networks are well connected, the role of hydraulic fracture is more dominant in the depletion process.

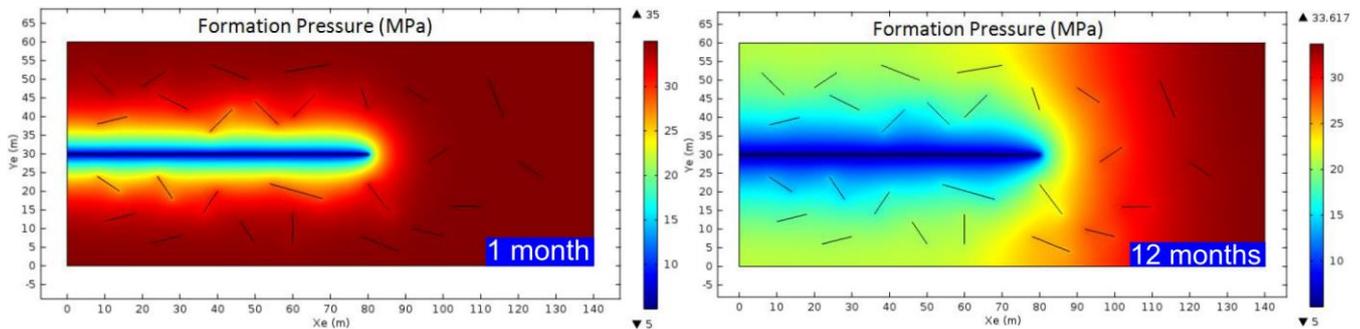

**Fig. 32 Pressure distribution in the SRV after 1 and 12 months of production with poorly connected fracture networks**

**Fig. 33** and **Fig. 34** show the impact of fracture connectivity and pressure dependent fracture permeability on gas production and production decline. The results pressure dependent fracture permeability does impact gas production and its decline trend if the fracture networks are well connected, as analyzed above. However, if the fractures are sparsely distributed in the SRV and not hydraulic coroneted to each other, it has negligible influence on well performance, if it has any at all. That's why the density of micro-seismic events generally do not correlate to production, because it does not reflect the connectivity of fracture networks and only these stimulated fractures that well connected to the main fracture can enhance the overall flow capacity in the SRV.

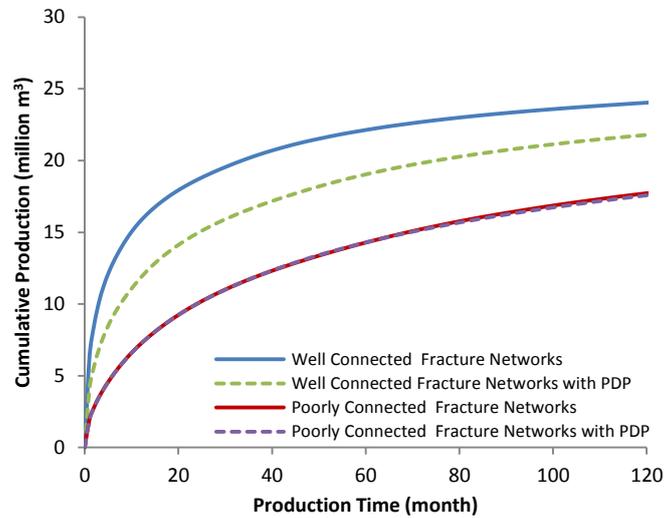

Fig. 33 Cumulative production with different fracture connectivity

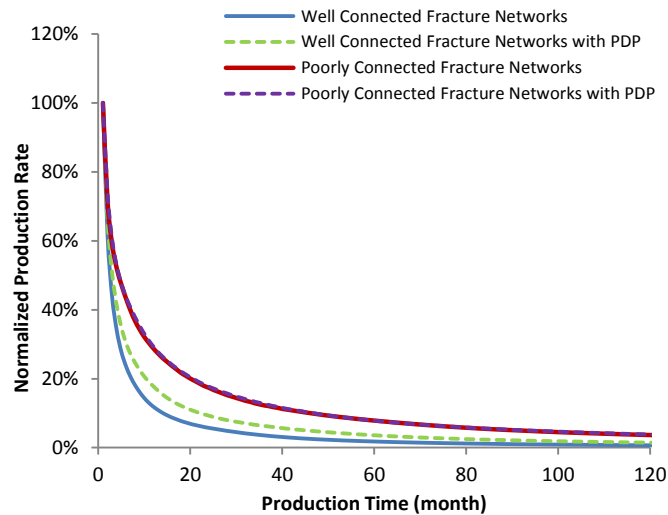

Fig. 34 Normalized production rate with different fracture connectivity

Here, it can be concluded that the sharp production decline trend in some shale plays cannot be caused purely by pressure dependent fracture permeability/conductivity. However, it is possible that rock creep or visco-elasto-plastic behaviors can reduce propped or un-propped fracture conductivity over time, which can cause steeper decline of production rate. Even though there are abundant literature on the changes of fracture permeability/conductivity as a function of pressure or confining stress, the time-dependent reduction in fracture permeability/conductivity is still limited, hence, will not be discussed further in this article.

Finally, the impact of initial reservoir pressure on the effects of pressure-dependent permeability is investigated. **Fig. 35** and **Fig. 36** show the impact of initial reservoir pressure. The results demonstrate that lower initial reservoir pressure leads to lower production and lower production decline rate. This is due to the fact that lower reservoir pressure result in less pressure drive for gas flow (under the same bottom-hole pressure) and less initial gas in place (less gas exists in the form of adsorption or compressed in pore space). In addition, the comparative reduction in fracture permeability is less when initial reservoir pressure is low (i.e., percentage of reduction in fracture permeability is lower if pressure declines from 20 MPa to 5 M Pa, instead of dropping from 35 MPa to 5 MPa, as reflected in Fig.26).

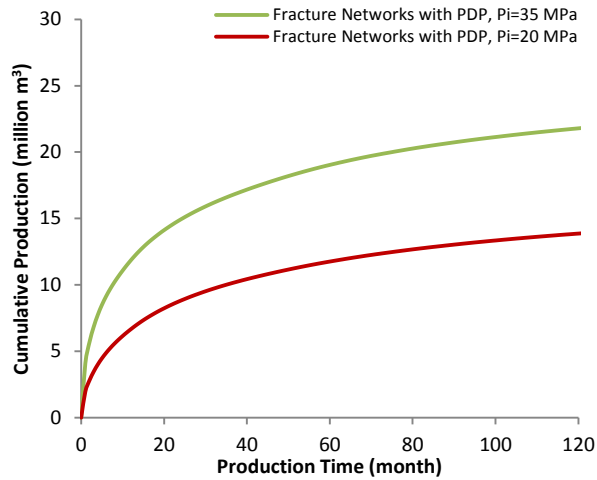

Fig. 35 Cumulative production with different initial reservoir pressure

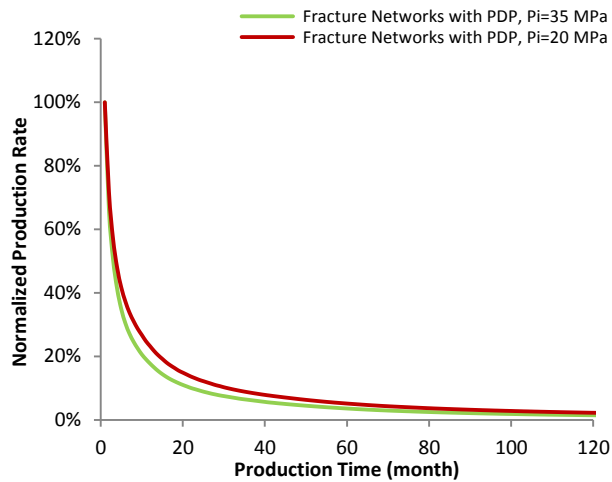

Fig. 36 Normalized production rate with different initial reservoir pressure

**Field Data Interpretation**

Valkó and Lee (2010) compiled monthly incremental production from selected Barnett Shale well groups, as shown in **Fig. 37**. The beginning of production from each well group starts from July of 2004, 2005 and 2006, respectively. From the graph, it can be observed that wells that are completed later, have higher initial production rate, but the production rate also declines faster. This can be explained by the fact that the improvement in technology and field operations makes well stimulation more efficient over time. With more hydraulic fracturing stages and better hydraulic fracture design, denser fracture networks and larger fracture surface area can be created, thus both initial production and production decline rate increase, which concurs well with above modeling results and analysis. Even though better stimulation design and larger effective fracture surface area lead to steeper production decline trend, they do enhance long-term recovery because of much higher production rate for both initial and later production stages.

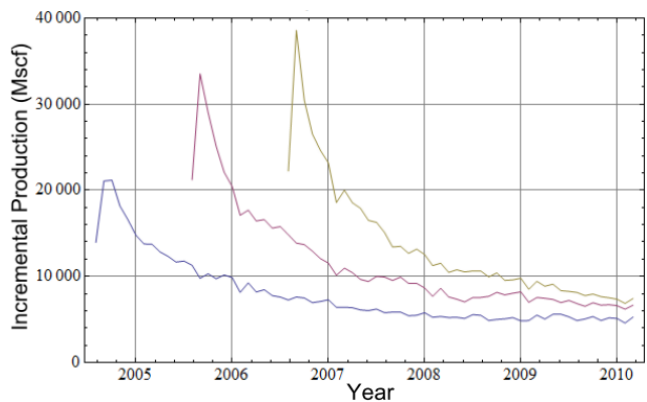

Fig. 37 Average monthly production increment from Barnet Shale wells (modified from Valkó and Lee 2010)

## Conclusions and Discussions

Predict post-fracture production rate and its decline trend in shale reservoirs with enough certainty is crucial for ultimate recovery estimation and economic decisions. Decline curve analysis is commonly used to determine EUR and in-place reserves are based on conventional reservoir performance assumptions. In unconventional reservoirs, these assumptions may not be valid, or they may be valid only so late in the reservoir life as to be little useful. Depending on the size of SRV, matrix permeability and the density, connectivity, conductivity of fracture networks, boundary dominated flow may not be apparent in the first a few years, making traditional decline curve analysis ineffective. In addition, empirical models for decline curve analysis cannot be used to analyze what factors influence the production and decline trend, which makes it difficult to guide completion designs or optimize field development.

In this article, we first presented some possible mechanisms that may impact shale gas production and its decline trend. Then, a proposed unified shale gas reservoir model, which is capable of modeling real gas transport and nano-flow mechanisms in fractured shale systems, is used to investigate what factors influence gas production and decline trend under coupled, multiphysics effects. The results of this study provide us a comprehensive understanding of the correlations between various factors and production in fractured shale gas wells. More in-depth knowledge regarding the effects of factors controlling the behavior of gas flow in stimulated wells can help us develop more reliable models to forecast shale gas production, decline trend and ultimate recovery. Conclusions reached from the analysis presented in this paper include

1. The existence of adsorption gas increases overall production and decreases the production decline rate. However, contrary to the common belief that adsorption gas becomes important only when the average pressure in the reservoir drops to a certain level during boundary dominated flow, the adsorption gas plays a more important role in early transient flow stages when the desorbed gas continues compensating pressure loss. And in the late time of pseudo-state flow, the further release of residual adsorption gas has much less influence on gas production.

2. Even though higher matrix permeability contributes to higher production rate, it also leads to steeper production decline trend in bounded SRV. And matrix permeability becomes less influential on gas production in the presence of abundant, conductive fracture networks.

3. The evolution of shale matrix apparent permeability during production can increase overall production, but has limited impact on production decline trend. If dense, well-connected fracture networks dominate the overall reservoir flow capacity, then the changes in matrix apparent permeability in nano-pores have a negligible impact on production and decline trend.

4. When SRV is surrounded by Non-SRV, production rate declines more gradually. The higher the matrix permeability, the lower the decline rate, because pressure loss in the SRV can be compensated more quickly by the surrounding Non-SRV

5. The density, conductivity and connectivity of fracture networks have dominant impacts on gas production and its decline trend, especially when matrix permeability is low. Fracture networks with dense, highly conductive and well-connected fractures, always lead to higher production rate and more rapid production decline.

6. It is a common misconception to think that the pressure (stress) dependent fracture permeability /conductivity reduce overall production and increase production decline rate. However, this study reveals that the pressure dependent fracture permeability does impair the overall production, but it actually reduce production decline rate, regardless of matrix permeability. This is due to the fact that the reduction of fracture conductivity happens soon after production start, instead of occurring gradually as reservoir pore pressure declines.

7. If the fractures are sparsely distributed in the SRV and not hydraulically well connected, then these fractures have negligible influence on well performance, if it has any at all. So does their conductivity evolutions.

8. The reason for much higher production decline rate in unconventional reservoirs than that in conventional reservoirs is due to the fact that more reservoir contact area that contributes to production in unconventional reservoirs. If gas is produced from a single vertical well, then the production decline trend is stable and gradual, regardless whether the reservoir itself is unconventional or not. So to maintain the overall production output in unconventional reservoir fields, new wells have to be drilled and completed at a much faster pace to compensate high production decline rate.

In general, more adsorption gas and the existence of Non-SRV can reduce the production decline rate, and denser, more conductive, better-connected fracture networks, with larger effective fracture surface areas in the SRV, always lead to higher production rate and steeper production decline trend. Once the drawdown begins, the increase of effective stress around conductive fractures result in degradation of fracture (propped and un-propped) conductivity, which decreases initial production rate, but also reduces production decline rate. So the reason for the cases of steep production decline in some ductile formations, such as Haynesville and Eagle Ford Shale, cannot be explained by lower matrix permeability, less abundant natural fractures or pressure (stress) dependent fracture permeability/conductivity. The possible mechanism of rapid production decline trend in ductile shale plays may result from rock creep and visco-elasto-plastic behaviors that reduce fracture conductivity over time (e.g., gradual embedment of proppant), instead of response to the increase of effects stress promptly. Quantitative characterization of the visco-elasto-plastic and ductility of tight shale reservoirs using indicators, such as rock mechanic properties, clay content/ maturation and organic content, is important to evaluate long-term production. Even though numerous literature have discussed the pressure (stress) dependent fracture permeability/conductivity, our understanding of the coupled physical process in time-dependent rock deformation and its effects on fracture conductivity is still limited, and more research efforts in this area are certainly needed.

## Acknowledgement

The author likes to express his sincere gratitude to Dr. Michael J. Economides, for his generous support and valuable guidance during the early stage of model development. Also thanks to the editors and several anonymous reviewers, whose insightful comments and suggestions significantly improved the quality and readability of this article.

## Appendix
### Formulations of Real Gas Transport and Nano-Flow Mechanisms in Fractured Shales

In this section, the mathematical formulas that implemented in our unified shale gas reservoir simulator for modeling and simulating gas production in fractured shale systems are presented and discussed. The fully coupled process in shale gas reservoirs involves fluid flow within the formation matrix and fractures, shale gas adsorption and desorption, real gas properties and in-situ stress that affected by local pressure. The mathematical framework for developing shale gas reservoir models presented here enables us to investigate the interactions among multi-physics phenomena in shale gas formations and examine their implications for flow in nano-porous medium, global well performance, production decline trend and ultimate recovery. All the equations presented in this section do not have any unit conversion parameters, any unit system can be applied (e.g., SI unit or Oil Field unit) as long as they are consistent.

### Fluid Flow and Material Balance in Shale Matrix and Fractures

Darcy's law with apparent permeability correction is used to model the gas flow rate within the shale matrix:

$$\boldsymbol{q}_g = -\frac{\mathbf{k}_a}{\mu_g} \cdot \nabla P \tag{A.1}$$

where $\boldsymbol{q}_g$ is the gas velocity vector, $\mu_g$ is the gas viscosity, $P$ is reservoir pressure and $\mathbf{k}_a$ is the matrix apparent permeability vector, which is pressure and temperature dependent. For isotropic and homogenous formations, $\mathbf{k}_a$ can be treated as scalar. The continuity equation within shale gas formation can be written as:

$$\frac{\partial m}{\partial t} + \nabla \cdot (\rho_g \boldsymbol{q}_g) = Q_m \tag{A.2}$$

where $m$ is the gas mass content per unit volume, $\rho_g$ is gas density, $Q_m$ is the source term and $t$ is the generic time. The gas mass content $m$ is obtained from two contributions:

$$m = \rho_g \phi_m + m_a \tag{A.3}$$

$\phi_m$ is matrix porosity and $\rho_g \phi_m$ is the free gas mass in the shale pore space per unit volume of formation, while $m_a$ is the adsorbed gas mass per unit volume of formation, which can be determined from the Langmuir isotherm (Langmuir, 1916):

$$m_a = \rho_m \rho_{gst} V_L \frac{P}{P + P_L} \tag{A.4}$$

where $\rho_m$ is the shale matrix density, $\rho_{gst}$ is the gas density at standard conditions, $V_L$ is the Langmuir volume and $P_L$ is Langmuir pressure. Combine Eq.(A.3) and Eq.(A.4), we can estimate the percentage of initial adsorption gas in-place:

$$\text{Percentage of initial adsorption gas in place} = \frac{\rho_m \rho_{gst} V_L P_i}{\rho_{g0} \phi_{m0}(P_i + P_L) + \rho_m \rho_{gst} V_L P_i} \times 100\% \tag{A.5}$$

where $P_i$ is initial reservoir pressure, $\rho_{g0}$ and $\phi_{m0}$ are gas density and matrix porosity at initial reservoir conditions. From Eq. (A.4), we can also estimate how much adsorption gas has been produced locally as pressure declines:

$$\text{Percentage of initial adsorption gas produced} = \frac{\frac{P_i}{P_i + P_L} - \frac{P}{P + P_L}}{\frac{P_i}{P_i + P_L}} \times 100\% \tag{A.6}$$

Discrete Fracture Network (DFN) with tangential derivatives can be used to define the flow along the interior boundary representing fractures within the porous medium. Flow behavior inside fracture is governed by:

$$\boldsymbol{q}_f = -\frac{\boldsymbol{k}_f}{\mu_g} \cdot d_f \nabla_T P \tag{A.7}$$

where $\boldsymbol{q_f}$ is the gas volumetric flow rate vector per unit length in the fracture, $\boldsymbol{k_f}$ is the fracture permeability tensor, $d_f$ is fracture width and $\nabla_T P$ is the pressure gradient tangent to the fracture surface. The expected reduction in fracture permeability caused by the increasing effective stress during production is can be described by power law relationship (Cho et al. 2013), here, we can use the following correlation:

$$k_f = k_{f,i} e^{-B\sigma_m} \tag{A.8}$$

where $k_f$ is the fracture permeability, $k_{f,i}$ is the fracture permeability at initial reservoir conditions, B is a fracture compaction parameter that can be determined from experimental data, and $\sigma_m$ is the mean effective stress, which is the mean total stress minus pore pressure. In our presented cases, the relationship between fracture permeability and local pore pressure is reflected in **Fig.26**. It assumes the fact that the conductivity of natural fracture is much smaller than that of hydraulic fracture and it is more sensitive to pressure/stress changes.

The continuity equation along the fracture reflects the generic material balance within the fracture:

$$d_f \frac{\partial \phi_f \rho_g}{\partial t} + \nabla_T \cdot (\rho_g \boldsymbol{q}_f) = d_f Q_f \tag{A.9}$$

where $\phi_f$ is the fracture porosity, and $Q_f$ is the mass source term, which can be calculated by adding the mass flow rate per unit volume from two fracture walls (left and right):

$$Q_f = Q_{left}^f + Q_{right}^f \tag{A.10 - a}$$

$$Q_{left}^f = -\frac{\boldsymbol{k}_a}{\mu_g} \frac{\partial P_{left}}{\partial \boldsymbol{n}_{left}} \tag{A.10 - b}$$

$$Q_{right}^f = -\frac{\boldsymbol{k}_a}{\mu_g} \frac{\partial P_{right}}{\partial \boldsymbol{n}_{right}} \tag{A.10 - c}$$

where $\boldsymbol{n}$ is the vector perpendicular to fracture surface.

**Real Gas Properties and Rock-Fluid Coupling**

The in-situ gas density is calculated according to the real gas law:

$$\rho_g = \frac{PM}{ZRT} \tag{A.11}$$

$T$ is the reservoir temperature, M is the gas average molecular weight and R is the universal gas constant. The Z-factor can be estimated by solving Equation of State (EOS) or using correlations for the gas mixtures. In this study, the Z-factor is calculated using an explicit correlation (Mahmoud, 2013) based on the pseudo-reduced pressure ($p_{pr}$) and pseudo-reduced temperature ($T_{pr}$):

$$Z = (0.702 e^{-2.5 T_{pr}})(p_{pr}^2) - (5.524 e^{-2.5 T_{pr}})(P_{pr}) + (0.044 T_{pr}^2 - 0.164 T_{pr} + 1.15) \tag{A.12}$$

The advantage of using explicit correlation is to avoid solving higher order equations respect to Z-factor, which leads to multiple solutions and increases computation efforts.

Gas viscosity is an intrinsic property of gas itself, which can be determined based on its compositions, pressure and temperature. Lee et al. Correlation (1966) is used to estimate the gas viscosity:

$$\mu_g = 10^{-4} K e^{X \rho_g^Y} \tag{A.13 - a}$$

and

$$K = \frac{(9.379 + 0.01607M)T^{1.5}}{209.2 + 19.26M + T} \tag{A.13-b}$$

$$X = 3.448 + \left(\frac{986.4}{T}\right) + 0.01009M \tag{A.13-c}$$

$$Y = 2.447 - 0.2224X \tag{A.13-d}$$

The porous medium is assumed to be perfectly elastic so that no plastic deformation occurs. The constitutive equation can be expressed in terms of effective stress ($\sigma_{ij}$,), strain ($\varepsilon_{ij}$), and pore pressure (*P*):

$$\sigma_{ij} = 2G\varepsilon_{ij} + \frac{2G\nu}{1-2\nu}\varepsilon_{kk}\delta_{i,j} - \alpha P\delta_{i,j} \tag{A.14}$$

where G is the shear modulus, $\nu$ is the Poisson's ratio, $\varepsilon_{kk}$ represents the volumetric strain, $\delta_{i,j}$ is the Kronecker delta defined as 1 for $i = j$ and 0 for $i \neq j$, and $\alpha$ is the Biot's effective stress coefficient. The strain-displacement relationship and equation of equilibrium are defined as:

$$\varepsilon_{ij} = \frac{1}{2}(u_{i,j} + u_{j,i}) \tag{A.15}$$

$$\sigma_{ij,j} + F_i = 0 \tag{A.16}$$

where $u_i$ and $F_i$ are the components of displacement and net body force in the *i*-direction. Combining Eq. (A.14)–(A.16), we have a modified Navier equation in terms of displacement under a combination of applied stress and pore pressure variations:

$$G\nabla u_i + \frac{G}{1-2\nu}u_{j,ii} - \alpha P\delta_{i,j} + F_i = 0 \tag{A.17}$$

**Shale Gas Apparent Permeability Correction**

Darcy's law cannot describe the actual gas behavior and transport phenomena in nano-porous media. In such nano-pore structure, fluid flow departs from the well-understood continuum regime, in favor of other mechanisms such as slip, transition, and free molecular conditions. The Knudsen number (Knudsen, 1909) is a dimensionless parameter that can be used to differentiate flow regimes in conduits at micro and nanoscale, for conduit with radius r, it can be estimated by

$$K_n = \frac{\mu_g Z}{Pr}\sqrt{\frac{\pi RT}{2M}} \tag{A.18}$$

**Table 1** shows how these different flow regimes, which correspond to specific flow mechanisms, can be classified by different ranges of $K_n$.

| $K_n$ | $0 - 10^{-3}$ | $10^{-3} - 10^{-1}$ | $10^{-1} - 10^1$ | $> 10^1$ |
|---|---|---|---|---|
| Flow Regime | Continuum | Slip | Transition | Free Molecular |

**Table 1. Fluid Flow Regimes Defined by Ranges of $K_n$ (Roy et al. 2003)**

The apparent permeability of shale matrix can be represented by the following general form:

$$k_a = k_\infty f(K_n) \tag{A.19}$$

where $k_\infty$ is the intrinsic permeability of the porous medium, which is defined as the permeability for a viscous, non-reacting ideal liquid, and it is determined by the nano-pore structure of porous medium itself. $f(K_n)$ is the correlation term that relates the matrix apparent permeability and intrinsic permeability. Sakhaee-Pour and Bryant (2012) developed correlations which are based lab experiments. They proposed a first-order permeability model in the slip regime and a polynomial form for the permeability enhancement in the transition regime using regression method:

$$f(K_n) = \begin{cases} 1 + \alpha_1 K_n & \text{Slip Regime} \\ 0.8453 + 5.4576K_n + 0.1633K_n^2 & \text{Transition Regime } 0.1 < K_n < 0.8 \end{cases} \tag{A.20}$$

where $\alpha_1$ is permeability enhancement coefficient in slip regime. To ensure the approximation of continuity of $f(K_n)$ at the boundary region of slip and transition regime, where no existing model available, $\alpha_1$ is set to be 4. Intuitively, Eqs. (A.18) and (A.20) predict net increases of $K_n$ and $f(K_n)$ with decreasing pore pressure, respectively. However, decreasing pore pressure can also lead to reduction in pore radii and, in turn, reduce the intrinsic permeability $k_\infty$. It can be seen from Eq. (A.19) that the evolution of matrix apparent permeability $k_a$ is determined by two combined mechanisms (i.e., the variations in $k_\infty$ and $f(K_n)$ during production).

Even though numerous studies (Lunati and Lee 2014; Sigal and Qin 2008; Shabro et al. 2012) have developed various theoretical models to account for the effects of gas diffusion in kerogen in permeability enhancement, but how to quantify these effects in complex, heterogeneous shale systems are still not well understood within scientific and industry community. In addition, all these proposed models do not account for the dynamic changes of stress and adsorption layer as pressure declines (as reflected in **Fig.3**) in a fractured system, which can significantly overestimate the effects of molecule diffusion on gas production. Recent study (Wang et al 2015b) also reveals that under most real shale-gas-reservoir conditions, gas adsorption and the non-Darcy flow are dominant mechanisms in affecting apparent permeability, while the effects of surface diffusion caused by adsorbed gas can be largely overlooked. So in this article, molecule diffusion will not be discussed in the following derivations for matrix apparent permeability.

Wang and Marongiu-Porcu (2015) conducted a comprehensive literature review on gas flow behavior in shale nano-pore space and proposed a unified matrix apparent permeability model, which bridges the effects of geomechanics, non-Darcy flow and gas adsorption layer into a single mode, by considering the microstructure changes in nano-pore space. In a general porous media, the loss in cross-section area is equivalent to the loss of porosity,

$$\frac{\phi}{\phi_0} = \frac{r^2}{r_0^2} \tag{A.21}$$

where variables with a subscript 0 correspond to their value at the reference state, which can be laboratory or initial reservoir conditions. Laboratory measurements by Dong et al. (2010) show that the relationship between porosity and stress follows a power law relationship, and can be expressed using the concept of mean effective stress $\sigma_m$ (Wang and Marongiu-Porcu 2015):

$$\phi = \phi_0 \left(\frac{\sigma_m}{\sigma_{m0}}\right)^{-C_\phi} \tag{A.22}$$

$C_\phi$ is a dimensionless material-specific constant that can be determined by lab experiments. For silty-shale samples, the values of $C_\phi$ range from 0.014 to 0.056 (Dong et al., 2010). Combining Eq. (A.21) and Eq. (A.22) leads to the relationship between pore radius and local stress

$$r = r_0 \left(\frac{\sigma_m}{\sigma_{m0}}\right)^{-0.5C_\phi} \tag{A.23}$$

When the adsorption layer is considered, the thickness of the gas adsorption layer, $\delta$, can be interpolated based on a Langmuir type functional relationship:

$$\delta = d_m \frac{P/P_L}{1 + P/P_L} \tag{A.24}$$

where $d_m$ is the average diameter of gas molecules residing on the pore surface. And the effective pore radius (A.23) can be modified as:

$$r = r_0 \left(\frac{\sigma_m}{\sigma_{m0}}\right)^{-0.5C_\phi} - d_m \frac{P/P_L}{1 + P/P_L} \tag{A.25}$$

The relationship between intrinsic permeability and pore radius can have the following relationship

$$\frac{k_\infty}{k_{\infty 0}} = \left(\frac{r}{r_0}\right)^\beta \tag{A.26}$$

where $\beta$ is the coefficient that define the sensitivity of permeability to the changes of pore radius. Different shale formations may have different nano-pore structure typology, which leads to different values of $\beta$. In this study, $\beta$ equals 2 by assuming the overall all intrinsic permeability resembles fluid flow in a capillary tube (Beskok and Karniadakis 1999). Combining Eq. (A.25), Eq. (A.26) and Eq. (A.19), we have the final expression of matrix apparent permeability:

$$k_a = k_{\infty 0} \frac{(r_0 \left(\frac{\sigma_m}{\sigma_{m0}}\right)^{-0.5C_\emptyset} - d_m \frac{P/P_L}{1+P/P_L})^\beta}{r_0^\beta} f(K_n) \qquad (A.27)$$

Compared with the apparent matrix permeability model proposed by Wang and Marongiu-Porcu (2015), Eq. (A.27) is a general formula for shale matrix apparent permeability, with additional parameter **β** to relate the changes in pore radius to the alterations in intrinsic permeability.

Based on the formulations introduced above, a unified shale gas reservoir simulator was constructed, with embedded discrete fracture networks and unstructured meshes. Newton-Raphson method (Ypma, 1995) and finite element analysis (Zienkiewicz and Taylor, 2005) are used to solve all the coupled equations numerically. The overall flow capacity inside the stimulated reservoir volume is controlled by both matrix apparent permeability and the properties (density, connectivity and conductivity) of discrete fracture network.

**Nomenclature**

| | |
|---|---|
| $B$ | = Parameter for fracture pressure dependent permeability, $Lt^2/m$, $1/Pa$ |
| $C_\emptyset$ | = Material constant for pressure dependent porosity |
| $d_f$ | = Fracture width, L, $m$ |
| $d_m$ | = Diameter of absorbed gas molecules, L, $m$ |
| $E$ | = Young's modulus, $m/Lt^2$, $Pa$ |
| $f(K_n)$ | = Non-Darcy flow correction term |
| $F_i$ | = Net body force along i direction, $m/Lt^2$, $Pa$ |
| $G$ | = Shear modulus, $m/Lt^2$, $Pa$ |
| $k_a$ | = Apparent gas permeability, $L^2$, $m^2$ |
| $\mathbf{k_a}$ | = Gas apparent permeability tensor, $L^2$, $m^2$ |
| $k_f$ | = Fracture permeability, $L^2$, $m^2$ |
| $\mathbf{k_f}$ | = Fracture permeability tensor, $L^2$, $m^2$ |
| $k_{f,i}$ | = Fracture permeability at initial reservoir conditions, $L^2$, $m^2$ |
| $k_\infty$ | = Matrix intrinsic permeability, $L^2$, $m^2$ |
| $k_{\infty 0}$ | = Matrix intrinsic permeability at reference conditions, $L^2$, $m^2$ |
| $K_n$ | = Knudsen number |
| $m$ | = Total gas content, $m/L^3$, $kg/m^3$ |
| $m_{ad}$ | = Gas adsorption mass per unit volume, $m/L^3$, $kg/m^3$ |
| $M$ | = Molecular weight, m/n, $kg/mol$ |
| $\mathbf{n}$ | = Normal vector to fracture surface |
| $P$ | = Reservoir pressure, $m/Lt^2$, $Pa$ |
| $P_L$ | = Langmuir pressure, $m/Lt^2$, $Pa$ |
| $P_{pr}$ | = Pseudo-reduced pressure |
| $\mathbf{q_f}$ | = Flow rate vector in the fracture per unit length, $L^2/t$, $m^3/s/m$ |
| $\mathbf{q_g}$ | = Velocity vector of gas phase, L/t, $m/s$ |
| $Q_f$ | = Mass source term in fracture, $m/L^3/t$, $kg/m^3/s$ |
| $Q_m$ | = Mass source term in matrix, $m/L^3/t$, $kg/m^3/s$ |
| $r$ | = Effective pore radius, L, $m$ |
| $r_0$ | = Effective pore radius at reference conditions, L, $m$ |
| $R$ | = Universal gas constant, $mL^2/t^2/n/T$, $8.3145 J/mol/K$ |
| $T$ | = Reservoir temperature, T, $K$ |
| $T_{pr}$ | = Pseudo-reduced temperature |
| $u_{i,j}$ | = Component of displacement, L, m |
| $V_L$ | = Langmuir volume, $L^3/m$, $m^3/kg$ |
| $Z$ | = Gas deviation factor |
| $\alpha$ | = Biot's coefficient |
| $\alpha_1$ | = permeability enhancement coefficient |
| $\beta$ | = Coefficient for pore radius dependent intrinsic permeability |
| $\delta$ | = Thickness of gas adsorption layer, L, $m$ |
| $\delta_{i,j}$ | = Kronecker delta |
| $\varepsilon_{ij}$ | = Elastic strain |
| $\varepsilon_{kk}$ | = Volumetric strain |
| $\mu_g$ | = Gas viscosity, m/Lt, Pa s |
| $\nu$ | = Poisson's ratio |
| $\rho_g$ | = Gas density, $m/L^3$, $kg/m^3$ |

| | | |
|---|---|---|
| $\rho_{gst}$ | = Gas density at standard condition, m/$L^3$, $kg/m^3$ | |
| $\rho_m$ | = Matrix density, m/$L^3$, $kg/m^3$ | |
| $\sigma_{ij}$ | = Effective stress, m/L$t^2$, $Pa$ | |
| $\sigma_m$ | = Mean effective stress, m/L$t^2$, $Pa$ | |
| $\sigma_{m0}$ | = Mean effective stress at reference conditions, m/L$t^2$, $Pa$ | |
| $\phi_f$ | = Fracture porosity | |
| $\phi_m$ | = Matrix in situ porosity | |
| $\phi_{m0}$ | = Matrix porosity at reference conditions | |